\documentclass[11pt,letterpaper]{JHEP3}
\usepackage{amssymb}
\usepackage{graphicx}
\def\beq{\begin{equation}}
\def\eeq{\end{equation}}
\newcommand{\bea}{\begin{eqnarray}}
\newcommand{\eea}{\end{eqnarray}}

\newcommand{\calH}{{\mathcal{H}}}
\newcommand{\Neff}{{N^*_\mathrm{eff}}}
\newcommand{\pd}{{\partial}}
\newcommand{\calA}{{\cal A}}
\newcommand{\Df}{{\lambda}}
\newcommand{\bk}{{\mathbf k}}
\newcommand{\etabar}{{\bar{\eta}_{ss}}}
\newcommand{\la}{\langle}
\newcommand{\ra}{\rangle}
\newcommand{\ud}{\mathrm d}
\newcommand{\mbe}{\mathbf e}
\newcommand{\fnl}{{f_{\mathrm{NL}}}}

\newcommand{\mx}{{\mathbf{x}}}
\newcommand{\calO}{{\mathcal{O}}}
\newcommand{\mpl}{M_{\mathrm{pl}}}
\newcommand{\dsbar}{{\la \dot \sigma \ra}}
\newcommand{\etath}{{\eta_\vartheta}}

\def\e{\begin{equation}}
\def\q{\end{equation}}
\def\m{\begin{eqnarray}}
\def\n{\end{eqnarray}}

\title{Multi-field Inflation with a Random Potential}
\author{S.-H. Henry Tye, Jiajun Xu and Yang Zhang \\
{\em Laboratory for Elementary Particle Physics,
 Cornell University, Ithaca, NY 14853, USA}}

\abstract{Motivated by the possibility of inflation in the cosmic landscape, which may be approximated by a complicated potential, we study the density perturbations in multi-field inflation with a random potential. The random potential causes the inflaton to undergo a Brownian-like motion with a drift in the $D$-dimensional field space, allowing entropic perturbation modes to continuously and randomly feed into the adiabatic mode. To quantify such an effect, we employ a stochastic approach to evaluate the two-point and three-point functions of primordial perturbations. We find that in the weakly random scenario where the stochastic scatterings are frequent but mild, the resulting power spectrum resembles that of the single field slow-roll case, with up to 2\% more red tilt. The strongly random scenario, in which the coarse-grained motion of the inflaton is significantly slowed down by the scatterings, leads to rich phenomenologies. The power spectrum exhibits primordial fluctuations on all angular scales. Such features may already be hiding in the error bars of observed CMB $TT$ (as well as $TE$ and $EE$) power spectrum and have been smoothed out by binning of data points. With more data coming in the future, we expect these features can be detected or falsified. On the other hand the tensor power spectrum itself is free of fluctuations and the tensor to scalar ratio is enhanced by the large ratio of the Brownian-like motion speed over the drift speed. In addition a large negative running of the power spectral index is possible.
Non-Gaussianity is generically suppressed by the growth of adiabatic perturbations on super-horizon scales, and is negligible in the weakly random scenario. However, non-Gaussianity can possibly  be enhanced by resonant effects in the strongly random scenario or arise from the entropic perturbations during the onset of (p)reheating if the background inflaton trajectory exhibits particular properties. The formalism developed in this paper can be applied to a wide class of multi-field inflation models including, e.g. the N-flation scenario. 
}

\begin{document}

\maketitle

\newpage

\section{Introduction}

The study of the inflationary universe is mainly on the slow-roll model with a single inflaton field. However, motivated by a variety of reasons, including those from superstring theory, multi-field inflation has received a lot of attention recently. An important reason to focus on the single field case is simplicity. A typical multi-field inflationary scenario is clearly very complicated and not predictive. There are special multi-field models which are easy to analyze while still predictive: models where the fields do not interact with each other during inflation were analyzed first in Ref.\cite{Starobinsky:1986fxa}.  A special case of this class of models is when all inflaton fields have the same potential, such as assisted inflation\cite{assisted}. While in more general cases, e.g. N-inflation\cite{Ninflation}, and similar scenarios\cite{Battefeld:2008py, Misra:2007cq}, each inflaton can have its own potential. 

If the inflaton fields couple to each other during inflation and reheating, the problem seems rather intractable. Motivated by the realization of inflation in superstring theory, we expect a multi-field inflationary scenario. This is particularly true if inflation takes place in the stringy cosmic landscape\cite{Kachru:2003aw}, where the number of moduli is probably as large as hundreds. Even if not all of them participate in inflation, it is likely that a large subset of moduli contribute directly or indirectly to the inflationary scenario. Their interaction can be rather complicated (maybe strongly interacting as well), generating a cosmic landscape that is random (aperiodic) in some directions and quasi-periodic (within certain field range) in others.

A realistic illustration could be the following potential for moduli $\rho_i$ and axions $\phi_i$\cite{axionf},
\begin{equation}
V_{T} (\rho_i, \phi_i) =  V(\rho_j) - \alpha_i \cos \left(\frac{\phi_i}{f_i} \right)  + \beta_{ij}\cos \left(\frac{\phi_i}{f_i} - \frac{\phi_j}{f_j}\right) + \dots  + U(\rho_i, \phi_i) ~.
\end{equation}
$f_i$'s are the axion decay constants and $\alpha_i =  M_i^4 e^{-S_{\mathrm{inst}}^{i}}  \gg \beta_{ij} $, where $M_i$ is a natural mass (say, the string) scale, and the $i$th instanton has action $S_{\mathrm{inst}}^{i}$. $V(\rho_j)$ contains the potential coming from the moduli and includes $D$-term contributions. Presumably, it contains the dominant contribution to the vacuum energy density in $V_{T}(\rho_i, \phi_i)$, while $U(\rho_i, \phi_i)$ expresses the couplings of the moduli and the axions. In the region where $V(\rho_j)$ is relatively flat or relatively localized, extensive inflation may take place, and the inflaton is represented by the collection of $\phi_i$ axions. With very small $\alpha_i$, the wavefunction is either not trapped (because binding in higher dimension typically requires a deeper attractive potential) or tunneling out of any classical vacuum site is very rapid. This potential is periodic except for the interaction term $U(\rho_i, \phi_i)$, which is perturbative in general, but it can also introduce relatively strong interactions at times. 
Here, $U(\rho_i, \phi_i)$ plays the role analogous to impurities in a crystal or doping in condensed matter physics. 

On a more general ground, the complicated scalar potential could be described using the random matrices\cite{Aazami:2005jf}. Generically, the potential probably looks rugged at short distances but smoother at coarse-grained level. We shall assume the randomness in the potential comes in different scales. Roughly speaking, we have in mind the following scenario: the small scale randomness of the potential causes the inflaton to undergo Brownian-like motion in the $D$-dimensional field space, allowing the entropic perturbations to feed into the adiabatic perturbations during inflation. The large scale behavior of the potential allows the inflaton to slowly move down in the landscape and eventually end the inflationary epoch. The ending can be brought about by the appearance of tachyons, fast tunneling to a low point in the potential, or a quick roll down when the inflaton suddenly reaches a steep slope.

It is important to point out\cite{Tye:2006tg} that the number of scalar fields relevant for inflation should be bigger than 2. Otherwise, the inflaton wavefunction will be trapped (i.e., localized) in the inflaton potential. If so, eternal inflation is unavoidable and inflation does not end. Our scenario implicitly assumes there is no eternal inflation and this assumption becomes more likely as $D$, the number of scalar fields participating in inflation, increases beyond 2. Here we have in mind a relatively large $D$ (say, dozens to hundreds).  

In slow-roll inflation, the large number $N_e$ of e-folds of inflation follows from the flatness of the inflaton potential. In general, this requires a fine tuning. Here, the large scale structure of the inflaton potential has to be reasonably flat too. However, the large scale flatness constraint is substantially relaxed compared to that in the slow-roll case, since the inflaton exhibits a random walk like motion and scatterings tend to slow down its overall motion. For this reason, we consider multi-field inflation in the landscape to be quite natural.

Motivated by this picture, it is reasonable to ask if one can make any predictions about such an inflationary scenario. Although the analysis is rather general under certain conditions, it helps to have a concrete example in mind. As an illustration, we could imagine an extended version of brane inflation, where a $D$3-brane is moving in the cosmic landscape. Its position during inflation includes both its position in the moduli space and its position within the 6-dimensional compactified bulk. Its motion includes classical rolling and scattering, which may result in percolation. Quantum fluctuation leads to density perturbation. Presumably,  quantum diffusion may be included as well. One may consider this scenario as a generalization of chain inflation\cite{chain}, which includes only repeated fast tunnelings, and related scenarios in Ref.\cite{Davoudiasl:2006ax}. 

Note that, in the absence of quantum fluctuation, no cosmological perturbation is generated in the classical but random trajectory of the inflaton. Even for a convoluted random looking trajectory, the classical path of the inflaton will be the same at every causal patch of the universe during inflation and so no cosmological perturbation is  generated in the absence of quantum effects. So the cosmological perturbation is entirely sourced by the quantum fluctuation. As is well known, in multi-field inflation, the quantum perturbation results in the adiabatic mode and the entropic modes of the cosmological perturbation. If the classical trajectory is a straight line in the field space, the entropic modes and the adiabatic mode remain decoupled, and the density perturbation comes entirely from the adiabatic mode. The randomness of the classical background allows for the random mixing of adiabatic and entropic modes, which is how the randomness of the classical background path enters into the cosmological perturbations, and how the resulting power spectrum incorporates the randomness of the inflaton potential. In general, a random potential will lead to a combination of classical percolation and quantum diffusion. However, knowing that the density perturbation is very small, i.e. $\delta \rho/\rho  \sim 10^{-5}$, it should be a very good approximation to treat the quantum diffusion as a quantum fluctuation around the classical (but complicated) background trajectory. In this case, the data will never be able to reveal individual bumps. Our analysis in this paper  is carried out in this approximation. A full analysis of quantum diffusion will be interesting but is beyond the scope of this paper. (Of course, one may take the point of view that, since $\delta \rho/\rho$ is so small, such an analysis may not be necessary.)

Here we would like to show that if the potential is random enough, some definitive statements about such a multi-field inflationary scenario are possible. In fact, the predictions can be quite specific and detail independent. 

\begin{itemize}
\item Even if the inflaton moves relatively fast, its drift velocity can be small, dictated by the slope of the potential over large field scales. Inflation naturally lasts for many e-folds.

\item One may easily recover the usual slow-roll, almost scale-invariant power spectrum, with a slight additional red tilt (up to 2\% when scatterings are frequent but mild) in the power spectrum index. 

\item The running of the power spectral index can be large, because the conversion of entropic modes into the adiabatic mode introduces additional scale dependence. 

\item The tensor to scalar ratio $r$ can be large. On one hand, it is suppressed by the amplified scalar power spectrum; on the other hand, it gets enhanced by a factor equal to the square of the ratio of the Brownian-like motion speed to the drift speed, which we expect to be much greater than one in the strongly random scenario. The net effect may still be an enhancement in $r$.

\item Non-Gaussianity is generically suppressed by the growth of the adiabatic mode after horizon exit due to the presence of extra light fields. However, it may get enhanced by resonant effects\cite{Chen:2008wn}. On the other hand, the large number of strongly coupled fields can enhanced the non-Gaussianity induced during the onset of (p)reheating.

\item We generically expect random fluctuations in the scalar power spectrum due to scatterings of the inflaton. Such fluctuation may appear with random amplitudes and periodicity. The precise shape of the power spectrum is not quite informative here, but the variance of such fluctuations is well determined by the underlying microscopic properties of the scalar potential (landscape). The currently observed $TT$ power spectrum, looking at the resolution of each multiple moments, is not smooth. The error bar of each data points (roughly $10\%$) is comparable to the fluctuation of data itself (for $100 \lesssim l \lesssim 800$)\cite{Nolta:2008ih}, so it is hard to claim any fluctuating features in the currently observed power spectrum. Furthermore, current analysis based on binned data points from a few tens of multiple moments may have smoothed out such fluctuations.  However, with more observational data coming in the future, we expect that analyzing the power spectrum at the resolution of single multiple moments is possible and the fluctuations in the power spectrum can certainly be detected or falsified if the error bars become substantially smaller than the primordial fluctuations. Furthermore, if such fluctuations are detected in the $TT$ power spectrum, one expects to see the same pattern in $TE$ and $EE$ correlations as well. 

\end{itemize}

The paper is organized as follows. The relevant properties of the multi-field inflationary scenario are reviewed in Section \ref{sec_rev}. In Section \ref{sec_infmf}, we consider the model where the randomness in the potential leads to a Brownian-like motion for the inflaton. That is, only classical percolation is included. The power spectrum is discussed in Section \ref{sec_pzeta} and the non-Gaussianity is discussed in Section \ref{sec_fnl}. In Section \ref{sec_dis}, we discuss the more general situation where the random scattering is also caused by quantum diffusion. We attempt to argue that the generic results obtained in Section \ref{sec_pzeta} and Section \ref{sec_fnl} may hold even in the presence of such quantum diffusion. Section \ref{sec_sum} contains the summary and some remarks. Some details are relegated to the two appendices.

\section{Review of the Multifield Inflation Formalism}
\label{sec_rev}

In this section, we review the multi-field inflation models for $D$ light canonical scalar fields. 
We start with the action
\begin{equation}\label{action}
S=\int \ud^{4}x\,\sqrt{-g}\, \left(\frac{R}{2} + X - V\left(\phi^{I}\right) \right)
\end{equation}
where we define $X \equiv - \frac{1}{2}\, g^{\mu\nu}\pd_{\mu}\phi^{I} \pd_{\nu}\phi_{I}$, with $I=1,2,\dots, D$. The 4 dimensional Friedmann-Lema\^itre-Robertson-Walker metric takes the form
\[
ds^{2}=g_{\mu\nu}dx^{\mu}dx^{\nu}=-dt^{2}+a(t)^{2}dx_{i}\, dx^{i}\:.\]
We have set $\mpl^{2}=8\pi G=1$.

In this paper, we will mainly consider the canonical kinetic term for simplicity. Generic kinetic terms could be a function of $X \equiv -\frac{1}{2}\, G_{IJ}(\phi) g^{\mu\nu}\pd_{\mu}\phi^{I} \pd_{\nu}\phi^J$\cite{Huang:2007hh}, or more generally (especially for the motion of D-branes) as shown in Ref.\cite{Langlois:2008qf}, the full action should be a function of $X^{IJ} \equiv -\frac{1}{2}\, g^{\mu\nu}\pd_\mu\phi^{I} \pd_\nu \phi^J$, and involves non-trivial terms like $X_{IJ} X^{IJ}$. To simplify the analysis, we consider only canonical kinetic terms. The formalism developed in this paper could be generalized to the scenarios with non-canonical kinetic terms. 

\subsection{The Background Equations}
\label{sbsec_bk}

The background equation of motion for the scalar field $\phi^I$ is, 
\begin{equation}\label{eq:eom_phi}
\ddot{\phi_{I}} + 3H\dot{\phi_I} + V_{I} =  0 ~, \quad  V_I \equiv \frac{\partial V}{\partial \phi^I} ~. 
\end{equation}
We shall consider both the short distance scale properties and the large distance scale properties of this multi-field inflaton potential $V(\phi)$. The coarse-grained potential will be denoted by $\bar V$, which leads to the drift velocity of the inflaton. The scattering by the short distance property of the potential will lead to Brownian-like motion.

The background Friedmann equations are
\[
3H^2=\rho= \frac{1}{2} \sum \dot{\phi_I}^2 + V ~, \quad \dot H = -X ~,
\]
In the presence of multi-fields, it is useful to define a set of orthonormal basis $\{ \mbe_n \}$ in field space. The first vector is defined to be along the direction of the background trajectory,
\[
\mbe_\sigma^I \equiv \frac{\dot \phi^I }{\dot \sigma}~, \quad \dot{\sigma}^2 \equiv 2X ~.
\]
The rest $D-1$ vectors are orthogonal to $\mbe_\sigma$ and they are collectively called $\mbe_s$. 
Among all the $\mbe_s$, it will prove useful later to choose one along the direction $\dot{\mbe}_\sigma$. This particular unit vector is denoted by $\mbe_\kappa$, and it points along the curvature radius of the background path. 

The background equation of motion can be projected to the $\mbe_\sigma$ direction, and we get
\begin{eqnarray}\label{eq:eom_sigma}
\ddot\sigma + 3H\dot\sigma + V_\sigma = 0 ~, \quad V_\sigma \equiv V_I \mbe_\sigma^I ~.
\end{eqnarray}
One should note that the basis $\{\mbe_n\}$ defined here does not point to fixed directions in the field space. They are defined on the background path of the inflaton, and their orientations are time dependent. 

\subsection{Linear Perturbations}
\label{sbsec_linpb}

In the longitudinal gauge, the metric perturbation can be written as
\[
ds^{2}=-(1-2\Phi)dt^{2}+a(t)^{2}(1+2\Phi)dx_{i}\, dx^{i} ~.
\]
In this particular gauge, the gauge invariant curvature perturbation $\zeta$ and scalar field perturbation $Q_I$ takes the form,
\begin{eqnarray}
\zeta = \Phi - H \frac{\delta \rho}{\dot \rho} ~, \quad Q_I = \delta\phi_I - \frac{\dot \phi_I}{H} \Phi ~.
\end{eqnarray}
$\zeta$ and $Q_I$ are related through the Einstein's equation, 
\begin{eqnarray}
\zeta + \frac{H}{{\dot\sigma}^2} \dot\phi_I Q^I  = \frac{-2\rho}{9(\rho+p)}\frac{k^2}{a^2 H^2}\Phi ~,
\end{eqnarray}
On super-horizon scales, the right hand side is negligible, and we have
\begin{eqnarray}
\zeta \simeq -\frac{H}{\dot\sigma} Q_\sigma ~, \quad Q_\sigma \equiv \mbe_\sigma^I Q_I ~.
\end{eqnarray}
According to Ref.\cite{Wands:2000dp}, $\zeta$ evolves with time on super-horizon scales, 
\begin{eqnarray}
\dot\zeta = -\frac{H}{\rho + p} \,\delta p_{\mathrm{nad}} ~, 
\quad \delta p_{\mathrm{nad}} = \delta p - \frac{\dot p}{\dot \rho} \delta\rho ~.
\end{eqnarray}
where $\delta p_{\mathrm{nad}}$ is the non-adiabatic pressure perturbation. 

For single field inflation, $\delta p_{\mathrm{nad}} = 0$, so $\zeta$ is conserved. However, in the case with  $D$ canonical scalar fields, on super-horizon scales where the spatial gradient term can be ignored, Ref.\cite{Gordon:2000hv} used a hydrodynamic approach and found that
%\footnote{See Appendix A for the factor 2}
\begin{eqnarray}
\dot \zeta = -\frac{2H}{\dot{\sigma}} \, \dot{\mbe}_\sigma^I \,Q_I  
\label{zdot}
\end{eqnarray}
One sees that $\zeta$ is generically not conserved when the background path makes a turn, i.e. $\dot{\mbe}_\sigma \ne 0$, and it is only sourced by entropic perturbations along the $\dot\mbe_\sigma$ direction. 

\subsection{The $\delta N$ Formalism}
\label{sec_dN}

The primordial curvature perturbation $\zeta(t, \mx)$ generated during multi-field inflation can also be described using the $\delta N$ formalism\cite{Sasaki:1995aw}, which not only reproduces the result in the previous section at linear order, but naturally allows us to go to second order perturbations as well. The main idea of the $\delta N$ formalism is that during inflation, separate Hubble volume evolves independently, and they differ by how much they have expanded relative to each other towards the uniform energy density slice at the end of inflation. The difference $\delta N$ is caused by the super-horizon quantum fluctuations of the inflaton field and is related to the large scale curvature perturbation $\zeta$ on the uniform energy density slice through 
\begin{equation}
\zeta = \delta N = N(\phi^*_I(\mathbf{x}), \phi_I^E (\phi_I^*)) -  N_e^F ~,  \label{deltaN}
\end{equation}
where $N_e^F$ is the number of e-folds on the spatially flat slice, and $N(\phi^*_I(\mathbf{x}), \phi_I^E (\phi_I^*))$ is the number of e-folds on the uniform energy density slice. $\phi^*_I(\mathbf{x})$ denotes the field configurations at the time $t^*$ of horizon crossing. $\phi^E_I$ denotes the field configurations at the end of inflation, and they generically depend on the initial $\phi^*_I$ in the multi-field case. For the slow-roll single field case, we have $\delta N = H \delta t$, which was used to evaluate the scalar density perturbation\cite{Guth:1982ec, Starobinsky:1982ee}. 
More generally, we can expand $\delta N$ as
\begin{equation}\label{dN_0}
\delta N = N_I \, Q^I + \frac{1}{2} N_{IJ} \left(  Q^I Q^J -  \la Q^I Q^J \ra \right) + \dots ~,
\end{equation}
where we have defined $N_I \equiv \partial N/\partial \phi^I$, and $N_{IJ} = \partial^2 N/ \partial\phi^I\partial \phi^J$. $Q^I$ is the perturbation of $\phi^I$ in the spatially flat gauge. For future reference, the perturbation along the $\mbe_\sigma$ direction is denoted by $Q_\sigma$ and perturbations along the $\mbe_s$ directions are denoted by $Q_s$. We have $N_\sigma\equiv N_I \mbe_\sigma^I$ and $N_s \equiv N_I \mbe_s^I$. 

For the single field case, the leading order term in the $\delta N$ expansion gives
\begin{equation}
\zeta = \frac{\ud N}{\ud\phi} \,Q = -\frac{H}{\dot\phi}\Big |_{t^*} \,Q ~.
\end{equation}
From the $\delta N$ perspective, since different Hubble volumes all follow the same trajectory, the only source of $\delta N$ is the initial value of $\phi$ they start with. $\delta N$ does not change once $\delta\phi$ get frozen on super-horizon scales, leading to the conservation of $\zeta$.

The non-conservation of $\zeta$ in the multi-field case can also be understood in the $\delta N$ formalism. 
We start by writing the e-folds along the unperturbed path as
\begin{eqnarray}
N = \int_{\sigma^*}^{\sigma_E(\phi^*_I)} \frac{H}{\dot\sigma} \,\ud \sigma ~, 
\end{eqnarray}
from which we perform the first order expansion,
\begin{eqnarray}\label{dN_1.0}
\delta N &=& -\frac{H}{\dot\sigma}\Big|_{t^*} (\pd_I \sigma^*) Q^I
+ \frac{H}{\dot\sigma}\Big|_{t_E} (\pd_I \sigma_E) Q^I \nonumber \\
&& + \int_{\sigma^*}^{\sigma_E} \frac{H}{\dot\sigma} \,\ud \left(Q^I \frac{\delta \sigma}{\delta\phi^I}\right)
- \int_{\sigma^*}^{\sigma_E} \frac{H}{\dot\sigma^2} \, Q^s \pd_s \dot\sigma \,\ud \sigma  ~.
\end{eqnarray}
The first term above comes from the shift in initial value $\sigma^*$, and it corresponds to the adiabatic perturbation in the single field case. The second term containing $\pd_I \sigma_E$ arises when the hyper-surface of the end of inflation is not orthogonal to the background inflaton path. The last two terms are non-local and depend on the complete inflaton trajectory after $t^*$. They capture the fact that under entropic perturbations, the inflaton follows a new trajectory nearby, and the new trajectory has different length (the third term) and also different speed (the fourth term), both contributing to $\delta N$. The last term only contains $Q^s$ perturbations, as it represents perturbations orthogonal to the integral path. 

The $\pd_I \sigma_E$ term is sub-leading when the inflaton executes Brownian-like motion, so we will ignore this term for a moment. In Section \ref{sbsec_fnlend}, when we look at the non-Gaussianity at the onset of (p)reheating, this term will be important. The non-local terms in Eq.(\ref{dN_1.0}) can be evaluated using a geometric approach (see Appendix \ref{app_a} for detailed derivations), and we get
\begin{eqnarray} \label{dN_1.1}
\delta N = -\frac{H}{\dot\sigma}\Big|_{t^*} Q^\sigma  
- \int_{t^*}^{t_E}  \ud t \, \frac{2H}{\dot{\sigma}} \dot{\mbe}^I_\sigma Q_I ~.
\end{eqnarray}
The $Q_\sigma$ perturbation has the same interpretation as in the single field case, it fluctuates tangent to the inflaton trajectory, leading to $\delta N$ between Hubble volumes that follow the same trajectory. However, there also exists fluctuations along the direction $\dot{\mbe}_\sigma$ (perpendicular to $\mbe_\sigma$), which kick the Hubble volume off the original expansion trajectory. 

The off-trajectory perturbations are exactly the source of the non-conservation of $\zeta$. In fact, using the relation $\zeta = \delta N$ together with Eq.(\ref{dN_1.1}), we get 
\[
\dot \zeta = -\frac{2H}{\dot{\sigma}} \, \dot{\mbe}_\sigma^I \,Q_I  
\]
which is the same as what one gets from the hydrodynamic analysis Eq.(\ref{zdot}).

By comparing the $\delta N$ expansion Eq.(\ref{dN_1.1}) with Eq.(\ref{dN_0}) at first order, we identify
\begin{eqnarray}
N_\sigma Q^\sigma &=& -\frac{H}{\dot\sigma} Q^\sigma \\
N_s\,Q^s |_{t^*} &=& -\int_{t^*}^{t_E}  \ud t \, \frac{2H}{\dot{\sigma}} \, \dot{\mbe}^I_\sigma \,Q_I ~. \label{NsQs}
\end{eqnarray}
It is also useful to write in components, 
\begin{equation}
N_\sigma|_{t^*} = -\frac{H}{\dot\sigma}~, \quad N_s|_{t^*}  = -\int_{t^*}^{t_E}  \ud t \, \frac{2H}{\dot{\sigma}} \, (\dot{\mbe}_\sigma \cdot \mbe^*_s) ~.
\end{equation}
We see that $N_s$ depends non-locally on the whole inflaton trajectory from $t^*$ to $t_E$, making it difficult to evaluate if the background inflaton trajectory makes several turns. 

We should also point out that unlike the adiabatic perturbations which get frozen on super-horizon scales, entropic perturbations could decay if the mass of entropic fields is higher than the Hubble scale. Especially, the moment when the inflaton path turns ($\dot\mbe_\sigma \ne 0$) could come much later than the moment when entropic perturbations are generated, and it is possible that by the time $\dot\mbe_\sigma \ne 0$, all the entropic perturbations have already decayed away. If that happens, we will get $N_s Q_s \approx 0$. 

In the context of a random potential, some of the entropic fields might be quite massive, while others are light compared to the Hubble scale. Only those light entropic fields contribute to the adiabatic mode and the number of such fields is $D$. It is entirely natural to have $D \gg 1$ in the context of string compactification. It is also possible that the mass of entropic fields fluctuates along the inflaton path. In such cases, if the mass does not remain larger than the Hubble scale for more than one e-fold, entropic perturbations may still survive. 

\section{Inflation with a Random Multi-Field Potential}
\label{sec_infmf}

One expects that the inflaton will follow a curved and random background trajectory in the $D$-dimensional field space. According to Eq.(\ref{NsQs}), every time the inflaton makes a turn ($\dot{\mbe}_\sigma \ne 0$), the perturbations along the $\dot\mbe_\sigma$ direction source $\delta N$. To determine $\delta N$, the evaluation of the integral in Eq.(\ref{NsQs}) is crucial. In principle, such an integration can be done given the knowledge of the scalar field potential and the inflaton trajectory in field space, and the result will be quite detail dependent. 

In this paper, we take an alternative point of view. We consider multi-field inflation with a random potential, and evaluate the power spectrum of the curvature perturbations generated by $D$ canonical scalar fields. We focus on the scenario where $D$ is large ($D$ could be 100 or larger); the scalar field interactions are complicated and can be strongly coupled, so the potential looks random (i.e., aperiodic). A typical scenario of such a scalar potential is the string landscape. In this scenario, the inflaton executes a Brownian-like motion in the $D$-dimensional field space even at the classical level. This type of random walk is different from the picture of stochastic inflation or eternal inflation, in which the random walk of the inflaton is driven by quantum fluctuations overwhelming the classical motion. Here the inflaton is mobile in the landscape. The random walk is due to classical scatterings in the random scalar potential and the resulting percolation. In our simplified scenario, the random classical scattering alone does not lead to density perturbation, which is sourced only by quantum fluctuation. However, the classical randomness of the potential does affect the mixing of adiabatic and entropic modes and thus will have impacts on the density perturbation by the end of inflation. 
In the next section, we shall focus on the scenario where the randomness comes from classical scatterings inside the random potential. The density perturbation comes from the quantum fluctuation on this classical background "random" inflaton trajectory. We will find that, when classical scatterings happen many times within one e-fold, the resulting power spectrum is relatively detail independent. In principle, a full analysis of quantum diffusion plus classical percolation will be interesting. However, since $\delta \rho/\rho \sim 10^{-5}$ is so small, such an analysis may not be necessary; that is, a perturbative treatment of the quantum effect on a classical background percolation is sufficient.

It is not hard to imagine the following scenario. Start with a smooth multi-field inflaton potential suitable for slow roll inflation. Now, let us slowly turn on many tiny localized bumps with different heights and sizes randomly distributed in that potential. (One may include other features such as small dips, steps, valleys and ridges too.) In the limit of zero height, we get back the original potential for slow roll. The question is what is the effect of the bumps on the power spectrum. If the inflaton encounters only one bump during its travel in one e-fold, then one expects a feature to show up in the power spectrum. The size and shape of the feature in the power spectrum obviously depends on the properties of the bump. This has been studied in the literature already. Note that, order of magnitude wise, there are about $400$ partial wave modes in the power spectrum per e-fold of inflation. Suppose that there is one data point per $20$ partial wave modes. So, if the inflaton encounters more than $20$ scatterings within one e-fold, we expect the distinct features due to the scatterings to smooth out in the data so one recovers a smooth observable power spectrum. It is in this limit that the statistical analysis presented here is fully applicable. As the number of scatterings decreases, our analysis becomes less reliable, but it also means that the likelihood of seeing some features in the power spectrum increases. We shall estimate the variance in the power spectrum due to such scatterings. 

How likely is such a potential in the cosmic landscape? As explained in the introduction, a potential with small bumps is actually quite likely, since the barriers in the axionic directions are expected to be exponentially small. Of course, a pure axionic potential is periodic, not random. However, its coupling to other moduli (and axions) will lead to additional interactions that interfere with the exact periodicities, like impurities in a crystal.  In fact, we hope that a detailed measurement of the power spectrum will tell us a lot about the particular patch of the landscape our universe has traveled over during the inflationary epoch.

The randomness of the potential can come in different scales. One can approximate the travel of the inflaton as a long term deterministic drift plus frequent stochastic scatterings on short time scales. At the coarse-grained level, the inflaton will try to drift down the potential, dictated by the slope of the potential over large field scales. At the same time, the small scale randomness causes the inflaton to bounce around many times within one efold, leading to fluctuations in $\dot\sigma$ on short time scales. We have in mind that these fluctuations are spaced by $\Delta N_e \ll 1$, i.e. the scatterings of inflation happen quite frequently within one e-fold. 

The coarse-grained deterministic motion down the potential results in a gradual change in the Hubble constant, and a small $\epsilon$ parameter, defined as
\begin{eqnarray}
\epsilon \equiv -\frac{\bar{\dot H}}{ H^2} ~,
\end{eqnarray}
where $\bar{\dot H}$ denotes the time average over the fluctuations within one e-fold.
As the inflaton moves down the potential, $\epsilon$ changes slowly, dictated by the $\eta$ parameter defined in the usual fashion by $\eta \equiv \dot\epsilon / (H \epsilon)$.

In the case with no or infrequent scatterings, we have $\epsilon = \dot\sigma^2/(2H^2M_{pl}^2)$. By analogy, we can define the coarse-grained speed $\dsbar$ at which the inflaton drifts down the potential by
\begin{equation} 
\epsilon = \frac{\dsbar^2}{2 \mpl^2 H^2}  \quad \Rightarrow \quad \dsbar \equiv H\sqrt{2\epsilon}\, \mpl ~. 
\end{equation}
In the following discussion, we shall always differentiate the drift speed $\dsbar$ from the Brownian-like motion speed $\dot\sigma$. Here, $\dot\sigma$ is the speed at which the inflaton travels between scatterings. We shall discuss two scenarios in the following sections. First, the weakly random scenario, where the drift term dominates over the scatterings. The inflaton undergoes frequent and mild scatterings, and after each scattering, $\mbe_\sigma$ changes direction slightly. We expect $\dsbar \approx \dot\sigma$. Second, we also have in mind the strongly random scenario where the scatterings change the direction of $\mbe_\sigma$ a lot. In this scenario, the stochastic scatterings dominate over the drift, and we generically expect $\dsbar \ll \dot\sigma$. 

To distinguish the weakly and strongly random scenario, we define a parameter $\chi$,
\begin{eqnarray}
\chi \equiv \frac{\dot\sigma^2}{\dsbar^2} ~,
\end{eqnarray}
$\chi \approx 1$ for the weakly random scenario and $\chi \gg 1$ for the strongly random scenario.  We note that 
$\dsbar$ measures the large distance properties of the potential while $\chi$ (or $\dot\sigma$) measures the short distance properties of the multi-field potential.

A common feature for both the weakly and strongly random scenarios is that entropic perturbation modes continuously feed into the adiabatic mode, resulting in the super-horizon evolution of the comoving curvature perturbations $\zeta$. 

The field space distance traveled by the inflaton in the down-hill motion may be estimated by relating $\epsilon$ to the overall tilt of the potential on large field scale $\nabla_\phi\bar{V}$. Approximately we have 
\[
\epsilon \approx \frac{\mpl^2}{2} \left|\frac{\nabla_\phi \bar{V}}{V}\right|^2  ~. 
\]
Using $\dsbar = \sqrt{2\epsilon}\, H \mpl$ yields a downhill field length of 
\begin{equation}
N_e \sqrt{2\epsilon}\, \mpl \approx \frac{\left|\nabla_\phi \bar{V}\right|}{V} N_e \mpl^2 ~. 
\end{equation}
However, on top of this deterministic down-hill motion, the inflaton also undergoes Brownian-like motion in the $D$ dimensional field space due to scatterings on short time scales. 

To get a concrete idea, we can model the random walk using the Langevin Equation
\begin{eqnarray}
\frac{\ud \vec{\phi}}{\ud t}  = -\frac{\nabla_\phi \bar{V}}{3H} + {\vec\xi}(t) = \vec{v}(\vec{\phi}) + {\vec\xi}(t) ~,
\end{eqnarray}
where all the vectors are defined in the field space. In general, $\left| \vec{v} \right| \ge \dsbar $. To simplify the model, we have identified the drift velocity
\[
\vec{v} =  -\frac{\nabla_\phi \bar{V}}{3H} ~, \quad \left| \vec{v} \right| = \dsbar ~, 
\]
and $\vec{\xi}$ is a random vector satisfying 
\begin{eqnarray}
\la \xi^I \ra &=& 0 ~, \\
\la \xi^I(t_1) \,\xi^J(t_2) \ra &\equiv& 2 D^{IJ}(\vec{\phi})\, \delta(t_1-t_2) ~. 
\label{DIJ}
\end{eqnarray}
with $D^{IJ}(\vec{\phi})$ is the diffusion tensor. Here we have assumed the white noise property for $\vec{\xi}$. This assumption is good if the scatterings take place at different sites in the landscape which presumably have uncorrelated properties and the typical time scale of the scattering itself is much less than the characteristic time between the scatterings.   

We can further work out $D^{IJ}$. We assume the characteristic time between scatterings is $\Delta t$, and the typical field space distance between two scattering sites is $\Delta\phi$. Upon each scattering, the change of the inflaton position in the field space is given by
\begin{equation}
\delta\vec{\phi} = \vec{\xi} \Delta t ~, \label{xi_delphi} 
\end{equation}
and $\delta\vec{\phi}$ can be decomposed along the direction of $\vec{v}$ and orthogonal to that,
\begin{eqnarray}
\delta\vec{\phi} = \delta\vec{\phi}_{\bot} + \delta\vec{\phi}_{\parallel} ~,
\end{eqnarray}
If we assume that the typical scattering angle is $\Theta$, in the weakly random case with $\Theta \ll 1$,
\begin{eqnarray}
\la \delta\phi^I_{\bot}(t_1) \delta\phi^J_{\bot}(t_2) \ra 
&\simeq& \frac{(\Delta\phi)^2}{D-1}  \delta^{IJ}_{\bot} \la \sin^2 \Theta\ra \, 
\delta \left( \frac{t_1}{\Delta t} - \frac{t_2}{\Delta t} \right)  \nonumber \\
&\sim& \frac{(\Delta\phi)^2 \Theta^2 \Delta t}{D-1} \delta^{IJ}_{\bot} \, \delta(t_1 - t_2)~, \\
\la \delta\phi^I_{\parallel}(t_1) \delta\phi^J_{\parallel}(t_2) \ra &\simeq& (\Delta\phi)^2  
\la (1-\cos \Theta)^2 \ra \, \mbe^I_{\mathbf v} \mbe^J_{\mathbf v} \,
\delta \left( \frac{t_1}{\Delta t} - \frac{t_2}{\Delta t} \right) \nonumber \\
&\sim&  (\Delta\phi)^2  \Theta^4 \Delta t \,  \mbe^I_{\mathbf v} \mbe^J_{\mathbf v} \, \delta(t_1 - t_2) ~. 
\end{eqnarray} 
Here we define $\delta^{IJ}_\bot = \delta^{IJ} - \mbe^I_{\mathbf v} \mbe^J_{\mathbf v}$, with $\mbe_{\mathbf v}$ the unit vector along the direction $\vec{v}$. We see that in the weakly random case, the longitudinal fluctuation is negligible. 

Since the drift velocity is also changing directions, we need to average over all directions of $\vec{v}$, this will change $\delta^{IJ}_\bot$ into $\delta^{IJ}(D-1)/D$. If we further
use the relation Eq.(\ref{xi_delphi}) and the definition of $D^{IJ}$ in Eq.(\ref{DIJ}), we get
\begin{eqnarray}
D^{IJ} \simeq \frac{(\Delta\phi)^2 \Theta^2}{2 D \Delta t} \delta^{IJ} ~, \quad \Theta \ll 1 ~. 
\end{eqnarray}
% In the weakly random scenario, $D^{IJ}$ is not isotropic, and the fact that $\delta^{IJ}_\bot$ appears means diffusion happens mainly in directions orthogonal to $\vec{v}$. 
If we introduce the diffusion constant $\Df$ defined as $D^{IJ} \equiv \Df \delta^{IJ}$, we have
\begin{eqnarray}
\lambda \simeq \frac{(\Delta\phi)^2 \Theta^2}{2 D \Delta t} ~, \quad \Theta \ll 1 ~. 
\end{eqnarray}

In the strong random scenario, when $\Theta$ randomly samples the interval $[0,\pi]$, we have
\begin{eqnarray}
\la \delta\phi^I_{\bot}(t_1) \delta\phi^J_{\bot}(t_2) \ra 
&\simeq& \frac{(\Delta\phi)^2  \Delta t}{D-1} \, \la\sin^2 \Theta\ra \, \delta^{IJ}_{\bot} \, \delta(t_1 - t_2)~, \\
\la \delta\phi^I_{\parallel}(t_1) \delta\phi^J_{\parallel}(t_2) \ra &\simeq& (\Delta\phi)^2 \Delta t 
\la \cos^2 \Theta \ra \, \mbe^I_{\mathbf v} \mbe^J_{\mathbf v} \, \delta(t_1 - t_2) ~. 
\end{eqnarray}
Note that the longitudinal fluctuation depends on $\cos(\Theta)$ instead of $(1-\cos\Theta)$. The reason is that when $\Theta \ll 1$, $\Delta\phi \cos\Theta \approx \Delta\phi$, and the fluctuation is given by $\Delta\phi \cos\Theta - \Delta\phi = (1-\cos\Theta)\Delta\phi$. However, when $\Theta$ takes all possible values from $0$ to $\pi$, $\Delta\phi \cos\Theta$ itself is the longitudinal fluctuation. 

Averaging over all directions of $\vec{v}$, we have $\delta^{IJ}_{\bot} \to \delta^{IJ}(D-1)/D$ and $\mbe^I_{\mathbf v} \mbe^J_{\mathbf v} \to \delta^{IJ}/D$. The diffusion tensor is 
\begin{eqnarray}
D^{IJ} \simeq \frac{(\Delta\phi)^2}{2D \Delta t} \delta^{IJ}
\left[ \la\sin^2 \Theta\ra  + \la \cos^2 \Theta \ra \right] \simeq  \frac{(\Delta\phi)^2}{2D \Delta t} \delta^{IJ} ~.
\end{eqnarray}
In summary we have the diffusion constant
\begin{eqnarray}
\Df &\simeq& \frac{(\Delta\phi)^2 \Theta^2}{2 D \Delta t} ~, \quad \Theta \ll 1 ~, \\
\Df &\simeq& \frac{(\Delta\phi)^2}{2D \Delta t} ~, \quad \Theta \sim 1 ~.
\end{eqnarray}

Equivalently, one may consider the time evolution of the probability density distribution $P(\vec{\phi}, t)$ of the inflaton position in the multi-dimensional field space, which is described by the Fokker-Planck equation,
\begin{equation}
\frac{\partial P}{\partial t}  = -\nabla \cdot \left[\vec{v} (\vec{\phi}) \,P \right] + \partial_I \partial_J 
\left[ D^{IJ}(\vec\phi) \,P \right] ~. 
\end{equation}

In the simple case, $\vec{v}$ is almost a constant vector and $D^{IJ} \equiv \Df\, \delta^{IJ}$ where $\Df$ is the diffusion constant. In the strong random case, with $\Theta$ random over $[0, \pi]$, we identify
\[
\Df = \frac{(\Delta\phi)^2}{4D \Delta t} ~. 
\]
With these simplifications, the Fokker-Planck equation can be easily solved.

Starting at $\vec\phi=0$ at $t=0$, the probability distribution of the inflaton at time $t=N_e/H$ is given by 
\[
P(\vec\phi,t) = (4 \pi \Df\, t)^{-\frac{D}{2}} \, 
\exp\left( -\frac{|{\vec\phi} - \vec{v}\,t|^2}{4 \Df\, t} \right)
= \left(4 \pi \Df \frac{N_e}{H} \right)^{-\frac{D}{2}} \,
\exp \left( -\frac{|{\vec\phi} - \vec{v}\,t |^2 H}{4 \Df N_e} \right)
\]
where the second expression is the probability distribution after $N_e$ e-folds of inflation. 
The amount of wandering explored by the inflaton is given by 
\[
\la \vec{\phi}^2 \ra = 2 D \Df \frac{N_e}{H}
\]
As $N_e$ increases, the inflaton explores a larger field space. Here the field space explored in $N_e$ number of e-folds is roughly 
\begin{eqnarray}
\dsbar t \, \la \vec{\phi}^2 \ra^{\frac{D-1}{2}} \sim  \dsbar (D \Df)^{\frac{D-1}{2}} 
\left(\frac{N_e}{H} \right)^{\frac{D+1}{2}} ~. \label{sample_patch}
\end{eqnarray}
Inflation ends when the inflaton reaches a big slope, or tunnels to a large dip in the potential. Another example is when the inflaton reaches a cliff (followed by tachyon condensation as in hybrid inflation or brane inflation). Knowing the structure of the landscape (or some region) will go a long way to see if sufficient number of e-folds is natural. We know that the field distance explored by the inflaton goes like $\sqrt{N_e}$, so naively many e-folds may be quite natural.  

We see that that the properties of the potential is captured in $\epsilon$, $D$, $\Df$ and $\chi$. In terms of independent parameters, we can trade $\Df$ for $\Delta t$. We should point out that the treatment of diffusion and drift in this section is very preliminary. In a realistic landscape, the drift velocity may not be a constant. The scattering angles many not follow a uniform distribution and the distribution may even vary over different parts of the landscape. The typical step size $\Delta \phi$ should also come with certain distributions and may be different in different directions. All these subtleties require a more careful treatment in the future. 

\section{The Power Spectrum}
\label{sec_pzeta}

As we have discussed in the previous section, the small scale randomness of the scalar potential causes the inflaton to undergo Brownian-like motion due to frequent scatterings within each e-fold. As a result, the background inflaton trajectory makes random turns.
To get an intuition on the behavior of the comoving curvature perturbation $\zeta$ in the presence of these random turns, we can first look at Eq.(\ref{zdot}). On the right hand side of Eq.(\ref{zdot}), $|\dot\mbe_\sigma|$ is now a random variable. Every time $\dot\mbe_\sigma \ne 0$, $\zeta$ gets a random jump. So the $D$-dimensional Brownian-like motion in the field space translates into the one dimensional random walk of the scalar quantity $\zeta$ through Eq.(\ref{zdot}). In fact, one can visualize this equation as the Langevin Equation for the one dimensional random walk\cite{Sethena:StatM}.

Using the $\delta N$ formalism, the power spectrum can be evaluated as
\begin{eqnarray}\label{dN_Pzeta}
P_\zeta = \frac{k^3}{2\pi^2} \la N_IQ^I N_J Q^J \ra 
= \frac{k^3}{2\pi^2} \left( \la N_\sigma Q^\sigma N_\sigma Q^\sigma \ra  + \la N_sQ^s N_{s'} Q^{s'} \ra  \right)
%\nonumber \\
%&=& \frac{H^2}{4\pi^2} \left( N_\sigma^2 + \la N_s N_{s'} \ra \delta^{ss'} \right) ~.
\end{eqnarray}
In the above equation, the summation over all the $s$ directions is implicit. If we further use 
\begin{eqnarray}
\la Q_\sigma Q_s \ra = 0 ~, && \quad \la Q_\sigma Q_\sigma \ra = \frac{H^2}{2k^3} ~, \nonumber \\
\la N_s Q^s N_{s'} Q^{s'} \ra &=& \la N_s N_{s'} \ra \la Q^s Q^{s'} \ra =  \la N_s N_{s'} \ra 
\frac{H^2}{2k^3} \delta_{ss'} ~,
\end{eqnarray}
then we get
\begin{eqnarray}
P_\zeta = \frac{H^2}{4\pi^2} \left( N_\sigma^2 + \la N_s N_{s'} \ra \delta^{ss'} \right) ~. 
\end{eqnarray}
The appearance of $\la N_s N_{s'}\ra$ in the power spectrum is a particular feature of our scenario, and deserves more clarification. As we have seen earlier, 
\begin{eqnarray}
N_s = \int_{t^*}^{t_E} \ud t \, \frac{2H}{\dot\sigma} \, \dot{\mbe}_\sigma^I \, \mbe^*_{s,I} ~.
\end{eqnarray}
When we calculate the power spectrum $P_\zeta$, we take the variance over different Hubble patches. Due to the initial quantum fluctuations, different Hubble patch follows slightly different paths, and they make turns of slightly different angles in the field space. We emphasize that the difference in $\dot{\mbe}_\sigma$ among different Hubble patches, is sourced by quantum fluctuations $Q_s$ frozen at $t_*$. In this sense, we can talk about $\la N_s N_{s'} \ra$, even if $N_s$ appears to be a classical quantity. Note that $N_s$ is always accompanied by $Q_s$ in the form $\la N_s Q^s N_{s'} Q^{s'} \ra$, and we have factored out the quantum fluctuation $\la Q_s Q_{s'}\ra$. Such an approximation amounts to assuming that $N_s$ and $Q_s$ are weakly correlated, which is probably true in our Brownian like motion. Since $Q_s$ is frozen at the horizon crossing time $t_*$, and $N_s$ received contributions every time inflaton scatters after $t_*$, we expect that after a few scatterings, the correlation between $Q_s$ and $N_s$ becomes very weak. However, it was exactly because of the entropic perturbations $Q_s$, that different Hubble patches will have slightly different $N_s$, and $\la N_sN_{s'} \ra$ will be part of the resulting power spectrum $P_\zeta$.  

As we have explained earlier, $\dot{\mbe}_\sigma \ne 0$ only when the inflaton makes a turn in the field space. We assume the characteristic frequency of such turns is $\Delta N_e$, i.e. there are $1/\Delta N_e$ turns per e-fold. We then introduce 
\begin{eqnarray} \label{ee}
\left\la \dot{\mbe}^I_\sigma(t_1) \dot{\mbe}^J_\sigma(t_2) \right\ra
= \frac{H}{\Delta N_e} \Theta^2(t_1) \delta_\bot^{IJ} \delta(t_1-t_2)  ~, 
\end{eqnarray}
where $\delta^\bot_{IJ} \equiv \delta_{IJ} - \mbe^\sigma_I \mbe^\sigma_J$. $\Theta^2$ is the variance of the angle turned in the field space, it is the variance among different Hubble patches when they make slightly different turns at roughly the same time. Typically we  expect $\Theta \lesssim 1$, since if different Hubble patches behaves too differently at each turn (large $\Theta$), $\delta N$ will be large by definition, and density perturbations can not remain small. 

Using the relation $\mbe^{*s}_I \,\mbe^{*s}_J = \delta^\bot_{IJ}$, we can write
\begin{eqnarray}
\la N_s N_{s'} \delta^{ss'} \ra &=& \frac{H^2 \vartheta}{\dsbar^2 \bar\chi}\Big|_{t*} \Neff ~,  \label{NsNs} \\
\vartheta(t) \equiv 4(D-1) \frac{\Theta^2(t)}{\Delta N_e} ~,\quad  &&
\Neff \equiv \int_{t^*}^{t_E} \frac{\epsilon^*\bar\chi^*}{\epsilon(t)\bar\chi(t)} \frac{\vartheta(t)}{\vartheta(t^*)} \, H \ud t ~. \label{neff}
\end{eqnarray}
In the above definition of $\Neff$, $\bar\chi$ is the time average of $\chi$ over one e-fold, it appears because we integrate over the history of inflation. $\chi$ it self may have fluctuations due to the Brownian-like motion, however $\bar\chi$ should be weakly time dependent. $\Neff$ denotes the effective number of e-folds from $t^*$ to $t_E$. In the limit when both $\vartheta$ and $\bar\chi$ are slowly varying, $\Neff$ approaches the actual number of e-folds $N_e^*$. The large scale observations of CMB temperature fluctuations today corresponds to $50 \lesssim N_e^* \lesssim 60$. 

The power spectrum now reads
\begin{eqnarray}
P_\zeta(k) = \left[ \frac{H^2}{8\pi^2 \epsilon \, \chi} + \frac{H^2}{8\pi^2 \epsilon \, \bar\chi} \vartheta \Neff \right] \Big|_{k=aH} ~. \label{Pzeta}
\end{eqnarray}
The first term comes from the pure adiabatic perturbation at horizon crossing. It depends on $\chi$ which may exhibit fluctuations due to the Brownian-like motion. The second term containing $\vartheta\Neff \sim D\Theta^2\Neff/\Delta N_e$ characterizes the amplification of the power spectrum by entropic perturbations after horizon crossing. The factor $D\Theta^2$ describes the number of dimensions the background random walk can turn to and the overall magnitude of the turning angle in field space. The factor $\Neff/\Delta N_e$($\gg 1$) is the effective number of steps the random walk takes towards the end of inflation. The $\vartheta\Neff$ should be smooth since it depends on $\bar\chi$. The full shape of the power spectrum could exhibit some oscillations on top of a smooth background. 

We will first look at the power spectrum in the weakly random case. The inflaton undergoes mild but still frequent scatterings while moving down the potential. The stochastic scatterings do not cause sharp changes in $\mbe_\sigma$, and we expect $\vartheta$ and $\epsilon$ to vary slowly too. Then we have $\chi \approx 1$, $\Neff \approx N_e^*$. The power spectrum now reduces to
\begin{eqnarray}
P_\zeta (k) &=& \frac{H^2}{8\pi^2 \epsilon} \left[1 + \vartheta N_e^* \right] \Big|_{k=aH}  ~. \label{Pzata_weak_ran}
\end{eqnarray} 
The spectral index can be calculated using the standard method, 
\begin{eqnarray}
n_s - 1 &=& \frac{\ud \ln P_\zeta}{\ud \ln k} \;\approx\; -2\epsilon -\eta - \frac{\vartheta}{1+N_e^*\vartheta} ~, 
\end{eqnarray}
We see that if $\vartheta N_e^* \gg 1$, $n_s - 1 = -2\epsilon - \eta - 1/N_e^*$. If $\vartheta N_e^* \ll 1$, then $n_s - 1 = -2\epsilon - \eta - \vartheta$. In either case, the correction to $n_s$ due to the multi-field effect is no greater than $-1/N_e^* \sim -0.02$. We conclude that in the weakly random case, the power spectrum Eq.(\ref{Pzata_weak_ran}) could easily resemble that of a single field slow roll model, and can easily agree with observational data. 

Second, we continue to look at the strongly random case, where both $\chi$ and $\vartheta$ vary with time. We expect that the long term drift speed $\dsbar$ is much smaller than the Brownian-like motion speed $\dot\sigma$, i.e. $\chi \gg 1$. During the Brownian-like motion, $\dot\sigma$ may fluctuate, leading to fluctuations in $\chi$, which will eventually impart oscillations on the pure adiabatic part of the power spectrum. Such oscillations in $P_\zeta$ could be different from those generated by sharp features in the potential\cite{stepmodel, Bean:2008na}, in that they may appear incoherent with random amplitudes and phases. Since the background Brownian-like motion undergoes many steps within one e-fold, the fluctuations in $P_\zeta$ should appear on all observable angular scales (with better chance of detection on small scales). The details of the oscillation in $P_\zeta$ cannot be fully captured by Eq.(\ref{Pzeta}) and numerical analyses are needed to calculate the evolution of $\zeta$ around horizon crossing carefully\cite{Bean:2008na}. However, the variance of the random oscillation might be estimated using Eq.(\ref{Pzeta}). To do that, we first averaged out the fluctuations in $P_\zeta$ and introduce the averaged power spectrum defined as
\begin{eqnarray}
\tilde{P}_\zeta(k) = \frac{H^2}{8\pi^2 \epsilon \, \bar{\chi}}  \left[1 + \vartheta \Neff \right] \Big|_{k=aH} ~, \label{Pzeta_aver} 
\end{eqnarray} 
The random oscillation about this average power spectrum can be characterized as
\begin{eqnarray}
\frac{\sqrt{\la (P_\zeta - \tilde{P}_\zeta)^2 \ra}}{\tilde{P}_\zeta} \sim \frac{\sqrt{\la (\chi -\bar\chi)^2 \ra}}{\bar\chi\, \vartheta \Neff} ~.
\end{eqnarray}
The variance of $\chi$ can be directly related to the short distance properties of the multi-field potential, which is an interesting question to explore further. 

The random oscillation of the power spectrum is a generic feature of our model, and may be falsifiable by future observations. If one looks at the un-binned $TT$ CMB power spectrum \cite{Nolta:2008ih}, one clearly sees that it fluctuates for about $10\%$ (for $100 \lesssim l \lesssim 800$). However, the error bars of each data point is also roughly $10 \%$, so it is hard to claim any fluctuating features in the $TT$ power spectrum based on current data. After binning a few tens of multiple moments into one data point, the current data analysis shows a relatively smooth $TT$ power spectrum over all angular scales. On the other hand, from $l\sim 100$ to $l \sim 1000$, there are about 2 or 3 efolds of inflation, so roughly there are 400 multiple moments per e-fold. If the inflaton encountered more than 20 scatterings per e-fold during inflation, then the current analysis based on the binned power spectrum will have smoothed out the fluctuations. In order to see such fluctuations in the power spectrum, it will be necessary to perform an analysis with the resolution of one or a few multiple moments. At current stage, such an analysis is still limited by statistics and systematics. With more data coming in the future, we expect to detect or falsify such primordial fluctuations in the power spectrum, if the error bars become substantially smaller than the primordial fluctuations. Furthermore, if such fluctuations are physical and appear in the $TT$ power spectrum, the same fluctuations should appear in $TE$ and $EE$ power spectra as well, which will provide consistency checks on their existence.  

The evaluation of the spectral index is subtle in this case. As we are mainly concerned with the overall tilt of the power spectrum, not that imparted by the fluctuations in $\chi$, the spectral index should be derived from the averaged power spectrum ${\tilde P}_\zeta$. 
\begin{eqnarray}
n_s - 1 = \frac{\ud \ln {\tilde P}_\zeta}{ \ud \ln k} 
&=& -2\epsilon -\eta -\eta_\chi - \frac{1 - \etath \Neff}{1+\vartheta \Neff} \,\vartheta ~, \\
\etath \equiv \frac{\dot{\vartheta}}{H \vartheta} ~, && \quad \eta_\chi \equiv \frac{\dot{\bar\chi}}{H\bar{\chi}} ~,
\quad 
\end{eqnarray}
We have introduced new parameters $\etath$ and $\eta_\chi$ to characterize the time dependence of $\vartheta$ and $\bar\chi$. $\etath$ and $\eta_\chi$ could be either positive or negative, unlike $\vartheta$ which is always positive. 

Last, we want to mention that since the tensor perturbations are decoupled from the scalar perturbations, the tensor mode power spectrum is unchanged, while the scalar power spectrum gets modified. The tensor to scalar ratio $r$ will be given by.
\begin{eqnarray}
P_T = 8 \left(\frac{H}{2\pi}\right)^2 ~, \quad
r = \frac{P_T}{{\tilde P}_\zeta} = \frac{16 \,\epsilon \, \bar{\chi}}{1+N_e^*\vartheta} ~. 
\end{eqnarray} 
We see that the conversion of entropic modes into the adiabatic mode suppresses $r$ by a factor of $1/(\vartheta\Neff)$. At the same time, $r$ also gets enhanced by $\bar\chi \gg 1$. Overall, $r$ gets modified by a factor of $\bar\chi/(\vartheta\Neff)$ compared to the single field slow-roll scenario. It is possible that this factor is greater than $1$, e.g. one may have $D\sim 100$, $\Delta N_e \sim 0.01$, $\vartheta \sim 10^4$, $\Neff \sim 50$, but $\chi$ can be as large as $10^6$, if the course-grained potential is flat enough. The current observation bound on the tensor mode $r \lesssim {\cal O}(0.1)$ gives, 
\begin{eqnarray}
\frac{\epsilon \, \bar{\chi}}{1+N_e^*\vartheta} \lesssim 0.01 \label{bound_r}
\end{eqnarray}

The analysis of the parameter space is more involved in the strongly random scenario. It is instructive to ignore $\eta_\chi$ for a moment (assuming $\bar\chi$ varies slowly with time) and focus only on the last term in $n_s$, which captures the effect of entropic perturbations. Below we will show two specific cases where some definite constraints can be drawn from observational data.

\begin{enumerate}

\item $|\etath \Neff | \gg 1$ and $\vartheta \Neff \gg 1$. The multifield effect on $n_s$ approaches
\begin{eqnarray}
\left| \frac{1 - \etath \Neff}{1+\vartheta \Neff}\right| \,\vartheta \; \sim \; |\etath| ~. \label{ns_bound1}
\end{eqnarray}
For self-consistency, $|\etath \Neff | \gg 1$ requires $|\etath| \gg 1/\Neff$. For $\etath < 0$, $\vartheta$ decreases with time and $\Neff \lesssim N_e^* \approx 60$, leading to $\etath \ll -0.02$. For $0 < \etath < 1$, we have $\Neff \gtrsim N_e^*$. We generically do not expect $\Neff$ to be much different from $N_e^*$, which ranges from $50$ to $60$. This leads to $\etath \gg 0.02$. Recent measurement of CMB\cite{Komatsu:2008hk} gives $n_s = 0.960^{+0.014}_{-0.013}$, so we could easily accommodate $-0.1 \lesssim \etath \lesssim 0.1$. With a large tensor to scalar ratio $r$, the upper-bound in $\etath$ can further be relaxed. 

\item $|\etath \Neff| \ll 1$ and $\vartheta\Neff \gg 1$, then we have 
\begin{equation}
\frac{1 - \etath \Neff}{1 + \vartheta \Neff} \,\vartheta \;\lesssim\; \frac{1}{\Neff} \label{ns_bound2}
\end{equation}
In this case, since $\etath \ll 1$, $\vartheta$ varies slowly and we expect $\Neff \sim N_e^*$. We may expect some running of the spectral index introduced by the $-1/\Neff$ term in $n_s$, 
\begin{eqnarray}
\frac{\ud n_s}{\ud \ln k} \approx -\frac{1}{\Neff^2}
\end{eqnarray}
The WMAP+ACBAR data\cite{Reichardt:2008ay} has shown that $\ud n_s /\ud \ln k =-0.037^{+0.023}_{-0.023}$ is consistent with current observations, which implies an lower bound $\Neff \gtrsim 10$. We thus require that the scalar potential on large field scale do not get tilted too much, so that $\epsilon$ does not increase too fast towards the end of inflation to give $\Neff \ll 10$. On the other hand, we also require entropic perturbations to dominate over adiabatic perturbations in the power spectrum to introduce such a scale dependence. If both conditions are met, our scenario could lead to detectable running of the spectral index. 
\end{enumerate}

In this section, we have seen that the background Brownian-like motion can impart two features on the power spectrum. The super-horizon evolution of $\zeta$ leads to a term proportional to $\vartheta \Neff$ in the power spectrum, which characterizes the conversion from entropic modes to the adiabatic mode. This contribution is continuous because the inflaton turns quite frequently in the field space within one e-fold. A second effect is that the small scale randomness of the potential leads to a ``refraction index'' $\chi$, which mainly affects the power spectrum during the epoch of horizon crossing. We have seen that in the weakly random regime ($\chi \approx 1$), the power spectrum resembles that of the single field slow-roll case, with up to 2\% more red tilt. While in the strong random regime ($\chi \gg 1$), the phenomenology could be quite rich. The tensor to scalar ratio can be enhanced and a large negative running of the spectral index may be observable. At the same time, the fluctuations in $\chi$ could impart (high frequency) random oscillations on the pure adiabatic part of the power spectrum. The details of such oscillations might not be picked out by the CMB data. However, the variance can be estimated based on the short distance properties of the scalar potential.

\section{Non-Gaussianity}
\label{sec_fnl}

Roughly speaking, possible sources of non-Gaussianity can come from three epochs:
during the inflationary stage, during the onset of (p)reheating and after the universe has entered into the hot big bang radiation era. 
In the following discussion, we will first consider the generation of non-Gaussianity during inflation using the $\delta N$ formalism. This non-Gaussianity turns out to be small because the integration over the inflationary epoch tends to wash out the fluctuating contributions, and the growth of $\zeta$ after horizon crossing suppresses the non-linear parameter $\fnl$ after proper normalization. 

Second, we will look at the  mechanism to generate non-Gaussianity at the onset of (p)reheating in our scenario, 
which was first studied in Ref.\cite{Lyth:2005qk, Dutta:2008if, Sasaki:2008uc}. This mechanism can take place in, for example, hybrid inflation, where tachyon instability ends inflation. The appearance of tachyons towards the end of inflation is quite natural in stringy inflationary scenarios. The abrupt ending of inflation is important for this scenario. Compared to Ref.\cite{Lyth:2005qk, Dutta:2008if, Sasaki:2008uc}, we may gain a factor of $D^2$ or $D^3$ due to the large number of fields participating, provided their effects act coherently.

 Although the best known mechanism to generate non-Gaussianity in the radiation epoch, namely, the curvaton model\cite{curvaton}, can be easily incorporated into our scenario, it is not a specific prediction of our model. So we will not discuss this possibility here. 

\subsection{Non-Gaussianity Generated During Inflation}
\label{sbsec_fnlinf}

The $\delta N$ formalism allows us to calculate the cubic order correlation of $\zeta$ in a relatively easier way\cite{Lyth:2005fi, Seery:2005gb}. The nonlinear parameter $\fnl$ is defined as
\begin{equation}
\left\langle \zeta(\mathbf{k}_{1})\zeta(\mathbf{k}_{2})\zeta(\mathbf{k}_{3})\right\rangle =(2\pi)^{7}\frac{\sum k_{i}^{3}}{\prod k_{i}^{3}}\left(\frac{3}{10} \fnl \right){\tilde P}_{\zeta}^{2}\,\delta(\mathbf{k}_{1}+\mathbf{k}_{2}+\mathbf{k}_{3}) ~.
\end{equation} 
Note that we normalize $\fnl$ in terms of the averaged power spectrum ${\tilde P}_{\zeta}$. 

If we write
\begin{equation}
\zeta = N_I \, Q^I + \frac{1}{2} N_{IJ} \left(  Q^I Q^J -  \la Q^I Q^J \ra \right) + \dots ~,
\end{equation}
following Ref.\cite{Seery:2005gb}, we can calculate the three-point function as
\begin{eqnarray}
\la \zeta(\bk_1)\zeta(\bk_2)\zeta(\bk_3)\ra &=& \la N_I N_J N_K \ra  \la Q^I(\bk_1) Q^J(\bk_2) Q^K(\bk_3)\ra  \label{fnl1} \\
&& +\; \frac{1}{2} \la N_I N_J N_{KL}\ra \la Q^I(\bk_1) Q^J(\bk_2) [Q^KQ^L](\bk_3) \ra + \textrm{perms}\dots ~. 
\label{fnl2}
\end{eqnarray}
The term in (\ref{fnl1}) contains the bispectrum, and the term (\ref{fnl2}) is the product to two power spectra. 
The bispectrum can be parametrized as
\begin{equation}
\la Q^I(\bk_1) Q^J(\bk_2) Q^K(\bk_3)\ra = (2\pi)^7 \,\delta\left(\sum \bk_i\right) \frac{\sum k^3_i}{\prod k_i^3} 
\left( \frac{H}{2\pi}\right)^4 \,{\cal A}^{IJK} ~,
\end{equation}
and we can calculate ${\cal A}^{IJK}$ using the standard method\cite{Maldacena:2002vr}, 
\begin{eqnarray} \label{inin}
\la Q^I(\bk_1) Q^J(\bk_2) Q^K(\bk_3)\ra = -i \int_{-\infty}^0 \ud t 
\left\la \left[ Q^I(\bk_1) Q^J(\bk_2) Q^K(\bk_3) , H_{\mathrm{int}}(t) \right] \right\ra  ~.
\end{eqnarray}
Here $H_{\mathrm{int}}$ is the cubic interaction Hamiltonian given by\cite{Langlois:2008qf, Gao:2008dt}
\begin{eqnarray}
H_{\mathrm{int}} &=& \int \ud^3 x  \,a^3
\left[ 
\frac{\dot\phi_K}{4H} \,\delta_{IJ} \left( \dot{Q}^I \dot{Q}^J Q^K + 2 \dot{Q}^I \pd^2 Q^J \pd^{-2} \dot{Q}^K \right)  
\right. \label{Hint_e} \\
&& + \left.
\left( \frac{\dot\phi_K}{4H} V_{IJ} + \frac{1}{6} V_{IJK} \right) Q^I Q^J Q^K
\right] \label{Hint_eta}
\end{eqnarray}
The interaction terms involving $V_{IJ}$ and $V_{IJK}$ are usually ignored in the single field slow-roll scenario, as they are higher order in slow-roll parameters. However, in the multifield case such assumptions might break down. In particular, $V_{IJK}$ could be large along the entropic directions. Furthermore, there exists small scale randomness that can cause $V_{IJ}$ to become momentarily large ($\gg H^2$) during the background Brownian-like motion. Therefore, we need to include both the $V_{IJ}$ and $V_{IJK}$ interactions when calculating the bispectrum. 

Following the standard method, the three point correlation function due to the interaction terms (\ref{Hint_e}) is, 
\begin{eqnarray}\label{A1}
{\cal A}_\epsilon^{IJK} =  \sum_{\mathrm{perm}} -i 
\int_{-\infty}^0 \ud \tau \, \frac{\dot\phi^I \delta^{JK}}{16 H} \, e^{iK\tau}
\left[k_1^2k_2^2(1-ik_3\tau) + 2k_1^2k_2^2(1-ik_2\tau)\right]+ \mathrm{c.c.}
\end{eqnarray}
The interaction terms (\ref{Hint_eta}) gives
\begin{eqnarray}\label{A2}
{\cal A}^{IJK}_{\epsilon \eta'} = \sum_{\mathrm{perm}}
-i \int_{-\infty}^0 \ud \tau\, \left(\frac{\dot\phi_K \eta_{IJ}}{32H} + \frac{V_{IJK}}{48H^2} \right)
\frac{e^{iK\tau}}{\tau^4} \prod (1-ik_i\tau) + c.c.
\end{eqnarray}
In the weakly random case, the above integrals can be estimated as
\begin{eqnarray}
{\cal A}_\epsilon^{IJK} \sim \frac{\dot\phi^I \delta^{JK}}{16 H} ~, \quad
{\cal A}^{IJK}_{\epsilon \eta'} \sim \frac{\dot\phi_K \eta_{IJ}}{32H} + \frac{V_{IJK}}{48H^2}  \label{A_est}
\end{eqnarray}
where we have assumed that $\dot\phi$, $\eta_{IJ}$ and $V_{IJK}$ are all slowly varying and we were not explicit about the momentum dependence. 

We point out that in the strongly random case, the careful evaluation of Eq.(\ref{A1}) and Eq.(\ref{A2}) is essential. Since $\eta_{IJ}$ and $V_{IJK}$ could fluctuate due to the background Brownian-like motion, there might be resonant enhancements in ${\cal A}^{IJK}_{\epsilon \eta'}$\cite{Chen:2008wn}. However, by the nature of randomness, the fluctuations in $\eta_{IJ}$ and $V_{IJK}$ must come at a range of frequencies with random amplitudes and relative phases, and the effects may not add up constructively and may just self average out. If it happens that there are a few frequencies that dominate the resonance, ${\cal A}^{IJK}_{\epsilon \eta'}$ might be enhanced. We leave such a possibility open and more detailed analysis is needed. 

Given the power spectrum and bispectrum of $Q^I$, the $\delta N$ formalism tells us how to reprocess $\la Q^3 \ra$ to get $\la \zeta^3 \ra$. Unlike the single field case, where $\zeta$ remains constant once leaving the horizon, here $\zeta$ will evolve even on super-horizon scales, and the evolution is exactly dictated by the e-fold derivatives $N_I$ and $N_{IJ}$ in the $\delta N$ formalism.  In order to estimate $\fnl$, we first need to know these derivatives precisely.  
From Section \ref{sec_dN}, we know that
\begin{eqnarray}
N_I &=& -\frac{H}{\dot\sigma} \mbe^\sigma_I 
- \int_{\sigma^*}^{\sigma_E}  \ud \sigma \, \frac{2H}{\dot{\sigma}} 
\frac{\ud \mbe^\sigma_I}{\ud \sigma} \label{NI} 
% \pd_I \sigma_E &=& - \int_{\sigma^*}^{\sigma_E} \ud \sigma \, 
% \sqrt{\frac{\epsilon^*\bar\chi^*}{\epsilon\bar\chi}}\, \frac{\ud \mbe^\sigma_I}{\ud \sigma} ~, \label{dse} 
\end{eqnarray}
We now continue to calculate $N_{IJ}$. Taking one more derivative in Eq.(\ref{NI}) gives
\begin{eqnarray}
N_{IJ} &=& -\pd_J\left(\frac{H}{\dot\sigma}\right) e^\sigma_I - \frac{H}{\dot\sigma}(\pd_J\mbe^\sigma_I)
+ \frac{2H}{\dot\sigma} \frac{\ud \mbe^\sigma_I}{\ud \sigma} (\pd_J\sigma^*) \nonumber \\
&& + \frac{2H}{\dot\sigma}\Big|_{t^*} \int_{\sigma^*}^{\sigma_E} \ud \sigma \, 
\sqrt{\frac{\epsilon^*\bar\chi^*}{\epsilon\bar\chi}}\, \frac{\ud \mbe^\sigma_J}{\ud \sigma}
\frac{\ud \mbe^\sigma_I}{\ud \sigma}
- 2\int_{\sigma^*}^{\sigma_E} \ud \sigma \,\mbe^s_J \, \pd_s \left(\frac{H}{\dot\sigma}\frac{\ud \mbe^\sigma_I}{\ud \sigma} \right) ~. 
\label{NIJ1}
\end{eqnarray}
The expressions for $N_I$ and $N_{IJ}$ simplify a lot if we rotate it into the $\mbe_\sigma$ and $\mbe_s$ directions. If we define 
$N_\alpha \equiv N_I \mbe_\alpha^I$, $N_{\alpha\beta} \equiv N_{IJ} \mbe^I_\alpha \mbe^J_\beta$, with $\alpha$ and $\beta$ representing either $\sigma$ or $s$, we have
\begin{eqnarray}
N_\sigma &=& -\frac{H}{\dot\sigma} \;=\; -\frac{1}{\sqrt{2\epsilon \chi}} \\
N_s &=& -\frac{2}{\sqrt{2\epsilon^*\bar\chi^*}} \int_{\sigma^*}^{\sigma_E}\ud\sigma\, \sqrt{\frac{\epsilon^*\bar\chi^*}{\epsilon\bar\chi}} \, \frac{\ud \mbe^\sigma_I}{\ud \sigma} \, \mbe^{*I}_s\\
N_{\sigma\sigma} &=& -\pd_\sigma \left (\frac{H}{\dot\sigma} \right) \;=\; \frac{\eta + \eta_\chi}{4\epsilon\chi} ~,  \\
N_{\sigma s} &=& -\pd_s \left(\frac{H}{\dot\sigma}\right) \;=\; 
\frac{1}{\sqrt{2\epsilon \chi}} \frac{\ud \mbe^\sigma_I}{\ud \sigma} \mbe^I_s ~, \\
N_{s s'} &=& \frac{8}{\sqrt{2\epsilon^*\bar\chi^*}}
\int_{\sigma^*}^{\sigma_E} \ud \sigma \, \sqrt{\frac{\epsilon^*\bar\chi^*}{\epsilon\bar\chi}}\, 
\frac{\ud \mbe^\sigma_J}{\ud \sigma} \frac{\ud \mbe^\sigma_I}{\ud \sigma} \mbe^{*J}_s \mbe^{*I}_{s'}  
+ \int_{\sigma^*}^{\sigma_E} \ud \sigma \, \frac{H}{\dot\sigma}\,
\frac{\eta_{IJ}}{\epsilon} \mbe^{*J}_s \mbe^{*I}_{s'}  ~,
\label{Nss}
\end{eqnarray}
where we have used Eq.(\ref{ds_dotsig}) in deriving $N_{ss'}$ above. 

We can now estimate the magnitude of non-Gaussianity. 
We first look at the bispectrum term (\ref{fnl1}), which leads to 
\begin{eqnarray}
\fnl = \frac{10}{3} \frac{\la N_I N_J N_K \ra \left(\calA_{\epsilon}^{IJK} + \calA_{\epsilon\eta'}^{IJK}\right)}{\la N_I N_J \delta^{IJ} \ra^2}  ~.
\end{eqnarray}

We first consider the weakly random case, the $\calA_\epsilon^{IJK}$ term contributes
\begin{eqnarray}\label{fnl_e}
\fnl \sim \frac{-5} {24 \la N_I N_J \delta^{IJ} \ra}
\sim - \frac{5\, \epsilon \bar\chi}{ 12(1+\vartheta \Neff)}  \sim -0.03\, r  ~.
\end{eqnarray}
This term turns out to be proportional to the tensor to scalar ratio $r$, and
using the constraint from the tensor mode Eq.(\ref{bound_r}), we see that the above $\fnl$ is negligible. 

The term $\calA_{\epsilon\eta'}^{IJK}$ contributes
\begin{eqnarray} \label{fnl_etap}
\fnl &\sim& \frac{5}{8}\frac{\la \eta_{IJ} N^I N^J \ra}{\la N_I N_J \delta^{IJ} \ra^2} 
+ \frac{5}{12}\frac{V_{IJK}}{H^2}\frac{\la N_I N_J N_K \ra }{\la N_I N_J \delta^{IJ} \ra^2}  \nonumber \\
&\sim& -\frac{5}{8} \left(\frac{\epsilon \bar\chi \,\eta_{\sigma\sigma}}{(\vartheta\Neff)^2}\frac{\bar\chi}{\chi} 
+ \frac{\epsilon \bar\chi \,\eta_{ss}}{\vartheta\Neff}\right)
- \frac{5}{12}\frac{\mpl}{H^2} \sqrt{2\epsilon\bar\chi} 
\left(\frac{V_{\sigma\sigma\sigma}}{(\vartheta\Neff)^2}\frac{\bar\chi^{3/2}}{\chi^{3/2}} + 
\frac{V_{\sigma s s}}{\vartheta\Neff} \frac{\bar\chi^{1/2}}{\chi^{1/2}}\right) ~. \nonumber \\
&\sim&  -\frac{\epsilon \bar\chi \,\eta_{ss}}{\vartheta\Neff} 
- \frac{\mpl}{H^2}\sqrt{2\epsilon\bar\chi} \frac{V_{\sigma s s}}{\vartheta\Neff} \frac{\bar\chi^{1/2}}{\chi^{1/2}}
\end{eqnarray}
In the above evaluation, we have taken $\Neff\vartheta \gg 1$ and kept only the leading order terms. We have also assumed that the background Brownian-like motion gives $\la N_s \ra = 0$ and $\la N_s N_{s'} N_{s''} \ra = 0$. Therefore, among all the $V_{IJK}$'s, only $V_{\sigma\sigma\sigma}$ and $V_{\sigma s s'}$ are left. 
We have defined new parameters
\[
\eta_{ss} \equiv \frac{\eta^{IJ} \delta^\bot_{IJ}}{D-1} ~, \quad  {V}_{\sigma s s} \equiv  \frac{V_{\sigma IJ} \delta_\bot^{IJ}}{D-1} ~. 
\]
We note that even if $\eta_{ss}$ and $V_{\sigma ss}$ could be locally large in the multifield scenario, their contributions to $\fnl$ are still suppressed by $1/(\vartheta\Neff)$. Especially the tensor mode constraint Eq.(\ref{bound_r}) further requires that $\epsilon \bar\chi \,\eta_{ss} / (\vartheta\Neff) \lesssim 0.01 \eta_{ss}$. We generically do not expect large $\fnl$ coming from Eq.(\ref{fnl_etap}). 

Again we should emphasis that the above results Eq.(\ref{fnl_e}) and Eq.(\ref{fnl_etap}) only apply to the weakly random case, where the scattering of inflaton is mild, although still being frequent. We still have entropic modes continuously feeding into the adiabatic mode. However, this conversion effect generically suppresses $\fnl$, as such an effect enhances $\la\zeta^2\ra$ and $\la \zeta^3 \ra$ equally by a factor of $\vartheta\Neff$. Therefore, $\fnl \sim \la\zeta^3\ra / \la\zeta^2\ra$ is generically suppressed by $1/(\vartheta \Neff)$. In the strongly random case, however, possible resonant effects might enhance $\fnl$, which is not captured in our results here. 

Second, we look at the term (\ref{fnl2}) arises from the second order expansion of $\delta N$ and it involves the product of two spectra, and gives
\begin{eqnarray}
\fnl &=& \frac{5}{6} \frac{N_I N_J N_{KL} \delta^{IK} \delta^{JL}}{(N_I N_J \delta^{IJ})^2} \nonumber \\
&\sim& \frac{\eta + \eta_\chi}{(\vartheta \Neff)^2} \frac{\bar\chi^2}{\chi^2}
+ \frac{2\vartheta}{(\vartheta \Neff)^2} \frac{\bar\chi}{\chi} + \frac{4\vartheta^2 N^*_1 + 4\vartheta\,\etabar\,\Neff N_2^*}{(\vartheta \Neff)^2}
%\frac{\eta + \eta_\chi + 2\vartheta + 4\vartheta^2 N^*_1 + 4\vartheta\,\etabar\,\Neff N_2^*}{(1+\vartheta \Neff)^2} 
\label{fnl3}
\end{eqnarray}
$\etabar$ is the time average of $\eta_{ss}$ over one e-fold. We have also defined new parameters
\[
N^*_1 \equiv \int_{t^*}^{t_E} \left(\frac{\epsilon^*\bar\chi^*}{\epsilon(t) \bar\chi(t)} \frac{\vartheta(t)}{\vartheta^*}\right)^2 
H \, \ud t ~, \quad
N^*_2 \equiv \int_{t^*}^{t_E} \frac{\epsilon^*\bar\chi^*}{\epsilon(t) \bar\chi(t)} \frac{\etabar(t)}{\etabar^*} \,H \, \ud t ~.
\]
Similar to $\Neff$, $N^*_1$ and $N^*_2$ are just weighed averaged number of e-folds. In the limit that $\epsilon$, $\chi$, $\vartheta$ and $\etabar$ are constants, we have $\Neff = N^*_1 = N^*_2$. 

Unlike $\fnl$ from the bispectrum, our result in Eq.(\ref{fnl3}) applies also to the strongly random case. We now look at how large this $\fnl$ could be. We assume $\Neff \sim N^*_1 \sim N^*_2$. More general cases can be analyzed similarly and the result remains the same qualitatively. We find that in Eq.(\ref{fnl3}), we can at most get one term that contributes $\fnl \sim 1/\Neff$, and another term that contributes $\fnl \sim \etabar/\vartheta$. We will discuss them separately below,

\begin{enumerate}
\item The $1/\Neff$ term.  As we have seen in the discussion on the power spectrum, scale invariance, i.e. the constraint on $\ud n_s/\ud \ln k$, requires $\Neff \gtrsim 10$, so this term contributes $\fnl < 1$. 

\item The $\etabar/\vartheta$ term. In order for the large scale entropic perturbations not to decay away, we need the mass of entropic fields to be well below the Hubble scale, which implies $\etabar \ll 1$. One may expect that a small $\vartheta$ can enhance $\fnl$; however, $\vartheta$ is also bounded from below. In order for the entropic perturbations to have an effect, we need $\vartheta\Neff \gg 1$, which leads to $\vartheta \gg 1/\Neff$. In the optimal case, we can take $\Neff \sim 50$, $\etabar \sim 0.1$, then $\vartheta \gg 0.01$. We may have a chance to get $\fnl \sim 1$ if we choose $\vartheta \sim 0.1$. However this is still too small to be detected by future experiments.  
\end{enumerate}

In this section, we have estimated the non-Gaussianity in our scenario. We conclude that $\fnl$ will generically be smaller than $\calO(1)$ in the weakly random case. The conversion of entropic modes into the adiabatic mode generically suppresses $\fnl$ by a factor of $1/(\vartheta\Neff)$. 
Large $\fnl$ could possibly arise in two situations. First, in the strongly random regime, $\fnl$ might be enhanced resonantly. Second, one may change the properties on the background random walk, e.g. by introducing the correlation of $\dot\mbe_\sigma$ at different times or at higher order, then one could expect to have $\fnl \gg 1$. We leave such possibilities for future study.

\subsection{Non-Gaussianity Generated During the Onset of (P)reheating}
\label{sbsec_fnlend}

Non-Gaussianity can be generated in hybrid inflationary models where inflation ends when a tachyonic field appears towards the end of inflationary epoch, which is quite natural in some scenarios inspired by string theory. In the string landscape with multi-fields, this can also be likely. Of course, both $\sigma$ and the tachyon field couple to other fields. The moment we are interested in is when the tachyon field first appears, when $\sigma=\sigma_E$. Let us focus on 2 specific terms in the $\delta N$ expansion,
\begin{equation}
\delta N =   -\frac{H}{\dot\sigma} Q^\sigma +  \frac{H}{\dot\sigma} \partial_s \sigma_E  Q^s 
+ \frac{1}{2} N_{ss'} \left( Q^s Q^{s'} - \la Q^s Q^{s'} \ra \right)  + \dots
\label{louis}
\end{equation}

Let $\varphi_s$ be the entropic field with positive mass (squared) and couple to $\sigma$, either directly or via the tachyon $T$. Let $U(\sigma, T, \varphi_s)$ be the potential. Since $\delta \varphi_s$ will decay outside the horizon, due to their masses, $m_s^2 \simeq \partial^2U/\partial \varphi_s^2$ (in the diagonal basis), we have
\[
\delta \varphi_s \simeq \frac{H}{2 \pi} \kappa_s ~, \quad \kappa_s \simeq e^{-m_s^2N_e/3H^2} ~.
\]
where $\kappa_s$ is the decay factor of the perturbation $\delta\varphi_s$. If we further assumes that  
$\pd_s \sigma_E$ is dominated by the moment when tachyon appears, we have
\[
N_s  = \frac{1}{\sqrt{2\epsilon}} \, \beta_s \Big|_{t_E} ~, \quad 
\beta_s \equiv \pd_s \sigma_E
\]
The non-Gaussianity generated is estimated in Ref.\cite{Dutta:2008if}. The $\varphi_s$ fields can be non-Gaussian due to their interactions and this non-Gaussianity can be transferred to $\zeta$ via the linear term in $\delta N$ (\ref{louis}). 
\begin{equation}
\la \zeta^3 \ra \sim \frac{1}{\epsilon^{3/2}}  \left\la (\beta_s \,\delta\varphi_s)^3 \right\ra \Big|_{t_E}
\end{equation}
This yields, according to Ref.\cite{Dutta:2008if}, 
\begin{equation}
\fnl \sim \frac{5}{9 \sqrt{2}} \frac{N_e \, \epsilon^{*2}}{H^2 \epsilon_E^{3/2}} \sum_{s\ge s' \ge s''}
(\beta_s\kappa_s) (\beta_{s'}\kappa_{s'}) (\beta_{s''}\kappa_{s''}) U_{ss's''}  ~, 
\end{equation}
where $\epsilon^*$ and $H$ are evaluated as horizon crossing. Since $\epsilon_E \sim 1$, we expect it to be much bigger than $\epsilon^*$. With typical $\kappa_s \ll 1$, we expect very small $\fnl$ generated here. Although the sum involves $D(D+1)(D+2)/6 \sim D^3$ terms, generically they may cancel among themselves to a large extent. It is also likely that the lightest $\varphi_s$ (with the largest $\kappa_s$) domintaes. Although we do not expect a large non-Gaussianity in general, the $U_{ss's''}$ terms may act coherently to make a big contribution to $\fnl$. This is entirely possible since the contribution does not involve a time integration which tends to wash out random or fluctuating effects.
 
The non-linear term also contributes to the non-Gaussianity. Because the non-linear term in $\delta N$ is local in spacetime, they will yield a contribution of the local form
\begin{equation}
\fnl \sim \frac{5}{3 \sqrt{2}} \frac{\epsilon^{*2}}{\epsilon_E^{3/2}} \sum_{s\ge s'}
(\beta_s\kappa_s^2) (\beta_{s'}\kappa_{s'}^2) \partial^2_{ss'} \sigma \Big|_{t_E}
\end{equation}
This term can go like $D^2$. Although we do not expect this local term to be too large generically, a large local non-Gaussianity is entirely possible.

\section{Discussions}
\label{sec_dis}

So far in this paper, we have considered a random potential where scatterings in the potential lead to inflaton motion resembling that of a random walk. We analyze the impact of this classical percolation property on the density perturbation sourced by quantum fluctuations. In a more realistic situation, we expect quantum diffusion effects on the motion of the inflaton as well. By this, we mean fast tunnelling, resonance scatterings as well as quantum hopping. Here, quantum hopping is hopping over a low barrier, due to the presence of the Gibbons-Hawking temperature $T=H/2 \pi$, so one may treat this as a thermal effect.  In this scenario, we assume that the inflaton is not trapped at any classically stable site that will lead to eternal inflation. Rather, the inflaton is mobile in the random potential and can move freely, via classical percolation or quantum diffusion. The argument for this property is based on the multi-field feature and discussed in Ref.\cite{Tye:2006tg}. There, the multi-field feature ($D \gg 1$) is crucial for the mobility of the inflaton. Here we see that the resulting magnitude of the density perturbation requires many scatterings during a single e-fold of inflation, consistent with the overall picture that emerges. 

Instead of the decay rate per unit volume $\Gamma$, it is convenient to introduce the tunneling probability in one Hubble time, the dimensionless parameter $\gamma$ given by $\gamma=\Gamma/H^4$, so $\gamma$ is a measure of the nucleation rate relative to the expansion rate of the universe. We are interested in situations where $\gamma > 1$. Let the probability of an arbitrary point remaining in a false de Sitter vacuum at time $t$ be $P(t)$. Then $P(t)$ behaves as $ P(t)\sim \exp ({-{4\pi\over 3}\gamma Ht})$. Thus the lifetime of the field in a false vacuum is estimated as 
\begin{equation}
t_F\simeq {3\over 4\pi H\gamma} = {3 H^3\over 4\pi \Gamma} 
\label{tf}
\end{equation}
If $\gamma \gg 1$, then $Ht_F \ll 1$.
Let us assume that inflation is dominated by repeated fast tunneling
events. Following Ref.\cite{Guth:1982ec, Starobinsky:1982ee}, or in the $\delta N$ formalism, we can easily estimate the
amplitude of density perturbation 
\begin{equation} 
\zeta = \delta N \simeq H \delta t \sim Ht_F 
\label{adp}
\end{equation} 
COBE normalization is $\zeta \sim
10^{-5}$ and thus $t_F\sim 10^{-5}H^{-1}$ which implies that there
are roughly $10^5$ steps during one e-fold. 
This is the key idea of chain inflation\cite{chain}.
We do like to note that an axionic potential with an exponentially small barrier height and a small decay constant do have the right feature for such a scenario.

Here, we assume that each site has a dominant decay channel for fast tunneling, so that the inflaton path is uniquely determined. Since $\gamma \gg 1$, the phase transition completes rapidly (that is, the universe percolates). So the inflaton path mimics a classical path. This puts a tight constraint on the inflaton potential. Presumably, this condition can be relaxed somewhat; that is, the intermediate steps can lead to different paths, though towards the end of inflation, the inflaton ends up in the same vacuum site in the potential. Such deviation in paths can lead to a new type of density perturbation, which must stay small to satisfy the observational bound. It will be very interesting to study this density perturbation and what it tells us about the inflaton potential.

Even with fast tunneling, a locally attractive site will trap the inflaton even for a little while. 
In high dimensions, it is harder to trap a wavefunction. So in some cases, the inflaton is not trapped by a local attractive site, but rather it scatters as a resonance over that site. In both cases, $\epsilon$ or $\dot \sigma$ will decrease and so $\zeta$ will increase. It is also likely that $\zeta$ fluctuates, with vales much bigger than its average $\sqrt{\la\zeta^2\ra} \sim 10^{-5}$. 
As long as such fluctuations happen at a scale too small to be picked up by the CMB data, (say, $10^3$ steps per e-fold), it will be difficult to detect this effect.

Such a scattering (or fast tunneling) from a meta-stable site A to a nearby site B can release a substantial fraction of the vacuum energy to the kinetic energy of the inflaton. Since the next step happens rapidly, there is no time for the Hubble expansion or the inflaton decay to dilute its kinetic energy, so next step scattering or tunneling can go to a site C with vacuum energy density comparable to that of A and above that of B. This means repeated scatterings and/or fast tunnelings do not have to go downhill. This leads to the Brownian-like motion. However, over some number of steps, the Hubble damping will take place and the inflaton will move downwards. This leads to the drift velocity. So at the coarse-grained scale, the resulting picture can easily mimic the slow-roll scenario.

Even though the lifetime is much shorter than the Hubble time, the motion of the scalar field does not really roll down along a smooth potential, but instead a bumpy potential. Usually the small bumps do not affect the amplitude of the power spectrum, but it may impart a large distinctive non-Gaussianity\cite{Chen:2008wn}. In our case the tunneling/percolation time is much shorter than the Hubble time, which means that the ``period'' of the bumps is very short and the non-Gaussian features from bumps may be too fluctuating to be picked up observationally. However, by the very nature of randomness in the potential, it is entirely possible that a big bump (or dip) in the potential or a particularly strong scattering can cause an observable feature in the CMB. This might be the origin of the feature at $l\sim 20$ in the WMAP data. Such fluctuations may appear in the power spectrum as a function of the wave number (or $l$). They are imprinted primordially, and should not be confused with the statistical fluctuations in data analysis.

One possible subtle issue we have not discussed in this paper is the effect of isocurvature perturbations after reheating. When the universe is reheated through the decay of a large number of scalar fields, the abundance of the created particles may depend on the initial value of the inflaton fields. The isocurvature perturbations can be characterized as the perturbation on the ratio of the particle number densities between different species, e.g. between photons and dark matter particles. Recent observations from WMAP+BAO+SN have put a stringent bound on the amount of isocurvature perturbations allowed in the primordial power spectrum\cite{Komatsu:2008hk}. The isocurvature power spectrum cannot exceed 6.7\% of its adiabatic counterpart. In our scenario, the power spectrum of adiabatic perturbation at the end of inflation is proportional to $\vartheta\Neff \sim (D-1)\Theta^2\Neff$, while we also have entropic perturbations by themselves with total power spectrum proportional to $(D-1)$, so the ratio of entropy perturbations over adiabatic perturbations is $1/(\Neff\Theta^2)$ which is about a few percent with $\Theta \lesssim 1$. The amount of isocurvature perturbations left after reheating could be well within the current observational bound. On the other hand, the reheating process may result in additional power in the curvature perturbation and possibly additional non-Gaussianity. This is certainly an interesting effect that deserves future study. 

\section{Summary and Remarks}
\label{sec_sum}

Motivated by the possibility of inflation in the cosmic landscape, which may be approximated by a complicated potential, we study the density perturbations in the multi-field inflationary scenario with a random potential. Here the randomness can come in different scales. It is a result of the complicated interactions among the moduli/axion fields in flux compactification. The small scale randomness of the potential causes the inflaton to undergo a Brownian-like motion in  the $D$-dimensional field space, allowing the entropic perturbations to continuously and randomly feed into the adiabatic perturbations. The large scale randomness allows the inflaton to slowly move down in the landscape and eventually end the inflationary epoch. 

To quantify the density perturbations in this scenario, we employ a stochastic approach to evaluate the two-point and three-point functions of primordial perturbations. We find that in the weakly random scenario where the stochastic scatterings are frequent but mild, the resulting power spectrum resembles that of the single field slow-roll case, with up to 2\% more red tilt. The strongly random scenario, in which the coarse-grained motion of the inflaton is significantly slowed down by the scatterings, leads to rich phenomenologies. The power spectrum exhibits such primordial fluctuations on all angular scales. Such features may already be hiding in the error bars of observed CMB $TT$ power spectrum and have been smoothed out by binning of data points. With more data coming in the future and improved systematics, we expect these features can be detected or falsified if one analyze the $TT$ power spectrum based on each single multiple moments. Furthermore, detection of such primordial fluctuations in the $TT$ power spectrum implies the same feature in the $TE$ and the $EE$ power spectrum as well. 
On the other hand the tensor power spectrum itself is free of such fluctuations and the tensor to scalar ratio is enhanced by the large ratio of the Brownian-like motion speed over the drift speed. In addition a large negative running of the power spectral index is possible.
Non-Gaussianity is generically suppressed by the growth of adiabatic perturbations on super-horizon scales, and is negligible in the weakly random scenario. However, non-Gaussianity can possibly  be enhanced by resonant effects in the strongly random scenario or arise from the entropic perturbations during the onset of (p)reheating if the background inflaton trajectory exhibits particular properties. 

The inclusion of quantum diffusion effect into this scenario remains to be worked out. The best guess made above is that it will not change the main consequences of our analysis.  We believe that inflation with a multi-field random potential can be systematically analyzed with solid predictions. The study in this paper provides a first step along this direction.

\vspace{1.5cm}

\noindent {\bf Acknowledgments}

\vspace{0.3cm}
We thank Xingang Chen, Eugene Lim, Louis Leblond, Tomohiro Matsuda, Liam McAllister, Enrico Pajer 
and Dmitry Podolsky for useful comments and discussions. 
This work is supported by the National Science Foundation under grant PHY-0355005.

\appendix

\section{An Explicit Derivation of $\delta N$ under Entropic Perturbations $Q_s$} 
\label{app_a}

\begin{figure}[t]
\centering
\includegraphics[width=9cm]{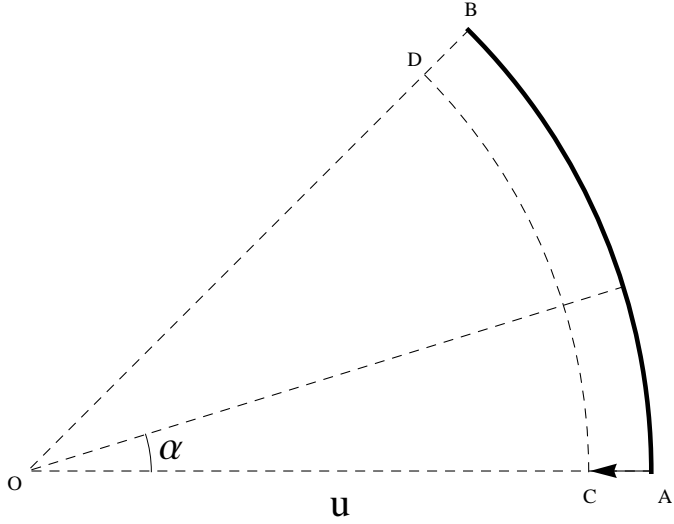}
\caption{Perturbation of the inflaton trajectory under $Q_s$. The solid line (AB) is the unperturbed trajectory, while the dashed line (CD) is the perturbed trajectory under the perturbation $Q_s$.}
\label{Fig:entropy}
\end{figure}

We now trying to calculated the perturbation of e-folds due to the entropic perturbation $Q_s$, by considering a short segment of the inflaton trajectory as shown in Fig.\ref{Fig:entropy}. The trajectory given by the dashed line (CD) is the perturbed trajectory from the solid line (AB) under $Q_s$. The two trajectories correspond to different number of e-folds due to their different length (curvature radius) and different 
$\dot\sigma$ along the trajectory. The number of e-folds along the unperturbed trajectory is given by
\begin{eqnarray}
N = \frac{H}{\dot\sigma} \Delta l ~,
\end{eqnarray} 
where $\Delta l = \dot\sigma \ud t$ is the length of the trajectory in the field space. Under the entropic perturbation $Q_s$, $\Delta l$ changes to \footnote{The sign convention is that the perturbation along the direction $\dot\mbe_\sigma$ corresponds to a shorter path.}
\begin{eqnarray}
\Delta l - \Delta\mbe^I_\sigma \, Q_I ~,
\end{eqnarray}
where $\Delta\mbe_\sigma$ is the change of the vector $\mbe_\sigma$ along the path. 

We also need the change of $\dot\sigma$ under $Q_s$. Since on super-horizon scales, $Q_s$ perturbations do not change the energy density $\pd_s\rho = 0$, and given that $\rho = \frac{1}{2}\dot\sigma^2 + V(\phi^I)$, we have
\begin{eqnarray}
\pd_s\dot\sigma = -\frac{V_s}{\dot\sigma} ~. 
\end{eqnarray}

On the other hand, using the background equation of motion Eq.(\ref{eq:eom_phi}) and Eq.(\ref{eq:eom_sigma}) one can easily show that 
\begin{eqnarray}
\dot\mbe_\sigma^I = \frac{-\dot\sigma V_I  +  \dot\phi^I V_\sigma}{\dot\sigma^2}  \quad 
\Rightarrow \quad  \dot\mbe_\sigma^I Q_I = -\frac{V_s}{\dot\sigma} Q^s ~,  \label{dotesig}
\end{eqnarray}
which can be written in components as
\begin{eqnarray}
-\frac{V_s}{\dot\sigma} = \mbe_s \cdot \dot{\mbe}_\sigma ~.
\end{eqnarray}
So we have 
\begin{eqnarray}
\pd_s\dot\sigma = \mbe_s \cdot \dot{\mbe}_\sigma ~. 
\end{eqnarray}

We can now evaluate $\delta N$ due to $Q_s$, we have
\begin{eqnarray}
N_s Q^s &=& -\frac{H}{\dot\sigma} \Delta\mbe^I_\sigma \, Q_I -\frac{H}{\dot\sigma^2} (\pd_s\dot\sigma) Q^s \Delta l \nonumber \\
&=& -\frac{H}{\dot\sigma} \Delta\mbe^I_\sigma \, Q_I -\frac{H}{\dot\sigma^2} \dot{\mbe}^I_\sigma  Q_I \Delta l ~, \label{dN_Qs}
\end{eqnarray}
The path length $\Delta l$ can be expressed as
\begin{eqnarray}
\Delta l = \frac{\dot \sigma}{|\dot\mbe_\sigma|} |\Delta\mbe_\sigma| ~, \label{Dl}
\end{eqnarray}
where ${\dot\sigma}/|\dot\mbe_\sigma|$ is the curvature radius, and $|\Delta\mbe_\sigma|$ is the angle spanned by $\Delta l$. 

Combing Eq.(\ref{dN_Qs}) and Eq.(\ref{Dl}), we get
\begin{eqnarray}
N_s Q^s = -\frac{H}{\dot\sigma} \Delta\mbe^I_\sigma \, Q_I 
-\frac{H}{\dot\sigma} \dot{\mbe}^I_\sigma  Q_I \frac{|\Delta\mbe_\sigma|}{|\dot\mbe_\sigma|} ~. 
\end{eqnarray}
The rate of change in $\delta N$ is given by
\begin{eqnarray}
\dot{\delta N} = \frac{N_s Q^s}{\Delta t} = \frac{-2H}{\dot\sigma} \dot{\mbe}^I_\sigma Q_I ~.
\end{eqnarray}
Using the relation that $\zeta = \delta N$, we have derived Eq.(\ref{zdot}) from the $\delta N$ formalism,
\[
\dot \zeta = -\frac{2 H}{\dot \sigma} \dot\mbe_\sigma^I Q_I ~.
\]
Our result here does not depend on slow-roll approximations. 

Another relation we want to derive is the variation of $\dot\mbe_\sigma$ along the entropic direction, which was used in deriving Eq.(\ref{Nss}). Starting from Eq.(\ref{dotesig}), we have
\[
\dot\mbe^\sigma_I = -\frac{V_I}{\dot\sigma} + \frac{V_J}{\dot\sigma}\mbe_\sigma^I\mbe_\sigma^J ~, 
\]
taking derivative along the $\mbe_s$ direction, we get
\begin{eqnarray}
\pd_s (\dot\mbe^\sigma_I) = -\frac{V_{Is}}{\dot\sigma} + \frac{V_{Js}}{\dot\sigma}\mbe_\sigma^I\mbe_\sigma^J
+ \frac{V_I}{{\dot\sigma}^2}\pd_s(\dot\sigma) - \frac{V_{J}\pd_s(\dot\sigma) }{{\dot\sigma}^2}\mbe_\sigma^I\mbe_\sigma^J ~.
\end{eqnarray}
Especially, if we project $\pd_s (\dot\mbe^\sigma_I)$ to the $\mbe_{s'}$ direction, we have
\begin{eqnarray} \label{ds_dotsig}
\mbe_{s'}^I\, \pd_s (\dot\mbe^\sigma_I) = -\frac{V_{ss'}}{\dot\sigma} + \frac{V_{s'}}{{\dot\sigma}^2} \pd_s(\dot\sigma) 
= -\frac{1}{\dot\sigma} \left[ V_{ss'} + (\mbe_{s'}\cdot \dot\mbe_\sigma) (\mbe_{s}\cdot \dot\mbe_\sigma) \right] ~.
\end{eqnarray}

\section{The Evolution of $Q_\sigma$ and $Q_s$ During Inflation} 
\label{app_b}
The evolution of the comoving curvature perturbation follows Eq.(\ref{zdot}) 
\begin{equation}
\zeta^\prime = -\frac{2\calH}{\sigma^\prime} |\mbe_\sigma^\prime| Q_\kappa ~, \quad \calH \equiv \frac{a^\prime}{a} ~. \label{eq:Rdot_c1}
\end{equation}
We have switched to conformal time $d\tau=dt/a$. We further introduce a transition matrix to quantify the time derivative of the field space basis, define
\begin{eqnarray}
\mbe_\alpha^\prime = \Theta_{\alpha\beta} \mbe_\beta ~, \quad \Theta_{\alpha\beta} = -\Theta_{\beta\alpha} ~. 
\end{eqnarray}
$\Theta$ is antisymmetric due to $(\mbe_m^I \mbe_{mI})^\prime = 0$. Since we have already chosen $\mbe_\kappa$ along $\mbe_\sigma^\prime$, we get 
\begin{equation}
\mbe_\sigma^\prime = \Theta_{\sigma\kappa} \mbe_\kappa ~, \quad \Theta_{\sigma \alpha} = 0 \;\; (\alpha \ne \kappa) ~.
\end{equation}
Now Eq.(\ref{eq:Rdot_c1}) becomes, 
\begin{equation}
\zeta^\prime = -\frac{2\calH}{\sigma^\prime} \Theta_{\sigma\kappa} Q_\kappa ~. \label{eq:Rdot_c2}
\end{equation}

The evolution equation of $Q_n$ can be written in a nice form if we introduce $v_\sigma \equiv a Q_\sigma$ and $v_s \equiv a Q_s$. 
\begin{eqnarray}
v_{\sigma}^{\prime\prime}+\left[k^{2}-\frac{z^{\prime\prime}}{z}\right]v_{\sigma}-2\left(\Theta_{\sigma\kappa}v_{\kappa}\right)^{\prime}-2\frac{z^{\prime}}{z}\Theta_{\sigma\kappa}v_{\kappa} &=& 0 ~, 
\quad z\equiv\frac{a\sigma^{\prime}}{\mathcal{H}} \label{eq:v1_simp} ~, \\
v_{s}^{\prime\prime} +\left[\left(k^{2}-\frac{a^{\prime\prime}}{a}\right)\delta_{sl} + \mu^2_{sl} \right]v_{l} 
-2\Theta_{sl}v_{l}^{\prime} &=& 0 ~, \quad(s \ne \sigma ,\, l \ne \sigma)  \label{eq:vs_simp}
\end{eqnarray}
where the effective mass matrix $\mu_{sl}$ for $v_s$ is given by
\begin{equation}
\mu^2_{sl} \equiv a^{2}V_{sl}-\Theta_{sl}^{\prime}+\Theta_{sk}\Theta_{kl} 
+  4\Theta_{\sigma\kappa}^{2}\delta_{\,s\kappa}\delta_{\,l\kappa} ~.
\end{equation}

The fact that $v_\sigma$ couples only to $v_\kappa$ can also be understood from Eq.(\ref{eq:Rdot_c2}), which can be rewritten as
\begin{eqnarray}
\zeta^\prime = -2\frac{\Theta_{\sigma\kappa}}{z}v_{\kappa} ~. \label{eq:Rdot_c3}
\end{eqnarray}
Since $\zeta = -v_\sigma / z$, we have
\begin{eqnarray*}
v_{\sigma}^{\prime} & = & z^{\prime}\zeta+\zeta^{\prime}z=z^{\prime}\zeta+2\Theta_{\sigma\kappa}v_{\kappa}\\
v_{\sigma}^{\prime\prime} & = & z^{\prime\prime}\mathcal{R}+z^{\prime}\mathcal{R}^{\prime}+2\left(\Theta_{\sigma\kappa}v_{\kappa}\right)^{\prime}
=\frac{z^{\prime\prime}}{z}v_{\sigma}+2\frac{z^{\prime}}{z}\Theta_{\sigma\kappa}v_{\kappa}+2\left(\Theta_{\sigma\kappa}v_{\kappa}\right)^{\prime}
\end{eqnarray*}
If we separate the solution to Eq.(\ref{eq:v1_simp}) into the homogeneous solution $v^0_{\sigma}$ and inhomogeneous solution $\hat{v}_\sigma$, Eq.(\ref{eq:Rdot_c3}) provides exactly $\hat{v}_\sigma$ on large scales $(k^2 \ll aH)$. Since only $v_\kappa$ appears in $\zeta^\prime$, we expect that $v_\sigma$ is sourced only by $v_\kappa$. The homogeneous solution $v^0_{\sigma}$ corresponds to the initial value of $\zeta$. 

In terms of the $\delta N$ expansion, $\delta N = N_\sigma Q^\sigma + N_\sigma Q^s$, $v^0_\sigma$ provides $Q^\sigma$ and integral effects of $v_\kappa$ provides $N_sQ^s$. In this sense, we claim that $Q^\sigma$ and $Q^s$ are uncorrelated here. Since the effective mass matrix $\mu^2_{sl}$ is not diagonal, in principle all the $Q^s$ perturbations are correlated. However, if the background inflaton trajectory is random, $\mu^2_{sl}$ is a random matrix, the random correlation between $Q_s$ will get washed out over time. So we claim that $Q_s$ are also uncorrelated among themselves.

\end{document}